\documentclass[12pt,preprint]{aastex}

\begin{document}
\title{The Environments around Long-Duration Gamma-Ray Burst Progenitors}

\author{Christopher L. Fryer\altaffilmark{1,2}, Gabriel
Rockefeller\altaffilmark{1,2}, and Patrick A. Young\altaffilmark{1,2}}

\altaffiltext{1}{Department of Physics, The University of Arizona,
Tucson, AZ 85721} 
\altaffiltext{2}{Theoretical Division, LANL, Los Alamos, NM 87545}

\begin{abstract}

Gamma-ray burst (GRB) afterglow observations have allowed us to
significantly constrain the engines producing these energetic
explosions.  Te redshift and position information provided by
these afterglows have already allowed us to limit the progenitors of
GRBs to only a few models.  The afterglows may also provide another
observation that can place further constraints on the GRB progenitor:
measurements telling us about the environments surrounding GRBs.
Current analyses of GRB afterglows suggest that roughly half of
long-duration gamma-ray bursts occur in surroundings with density
profiles that are uniform.  We study the constraints placed by 
this observation on both the classic ``collapsar'' massive star
progenitor and its relative, the ``helium-merger'' progenitor.  We
study several aspects of wind mass-loss and find that our
modifications to the standard Wolf-Rayet mass-loss paradigm are not
sufficient to produce constant density profiles.  Although this does
not rule out the standard ``collapsar'' progenitor, it does suggest a
deficiency with this model.  We then focus on the He-merger models and
find that such progenitors can fit this particular constraint well.
We show how detailed observations can not only determine the correct
progenitor for GRBs, but also allow us to study binary evolution
physics.

\end{abstract}


\section{Introduction}

Since the concurrent and cospatial observation of supernova 1998bw
with gamma-ray burst 980425~\citep{Gal98}, evidence that long-duration
gamma-ray bursts (GRBs) arise from the explosion of massive stars has
continued to grow.  This supporting evidence includes the association
of long-duration GRBs with star-forming galaxies \citep{Fru02,Vre01}
and the detection of 1998bw-like ``bumps'' found in GRB afterglows
\citep{Blo03}.  The connection between long-duration GRBs and massive
stars was sealed by the paired observation of SN 2003dh and GRB 030329
\citep{Hjo03,Sta03}.  This association tells us that the progenitors
of long-duration GRBs are almost certainly massive stars.  Both theory
and observation further constrain the nature of these massive star GRB
progenitors.  A successful GRB engine must produce beamed jets of
relativistic matter with Lorentz factors in excess of 100 (Frail et
al. 2001).  To reach and maintain the high Lorentz factors observed in
GRB ejecta, the jet must not sweep up much mass and the jet engine
must be active for the entire time during which the jet pushes through
the star, leading theorists to argue that the massive star must lose
its hydrogen mantle before exploding (Woosley 1993; Zhang, Woosley, \&
McFadyen 2003). Such a star would have been a Wolf-Rayet star before
collapse and its resultant supernova explosion should have matched
type Ib/c supernovae.  Spectral observations of SN 2003dh have
confirmed that the progenitor was a Wolf-Rayet progenitor
\citep{Hjo03,Sta03}.

Wolf-Rayet stars are known to have strong winds with mass outflow
rates beyond $10^{-5} {\rm M_\odot y^{-1}}$ and wind velocities above
$1000 {\rm km\,s^{-1}}$.  Stellar winds sweep out the environment
surrounding the star and, depending upon the density of the ambient
medium, these wind-blown bubbles can extend beyond 10\,pc
\citep{Dwa05}.  The density profile for a free-streaming wind is
simply determined by mass conservation: $\dot{M}_{\rm wind}=4 \pi r^2
\dot{r} \rho(r) \rightarrow \rho(r) = \dot{M}_{\rm wind}/4 \pi r^2
v_{\rm wind} \propto r^{-2}$ where $\dot{M}_{\rm wind}$ is the mass
loss rate of the wind (assumed to be constant) and $\rho, v_{\rm wind}$ are
the density and velocity (or $\dot{r}$) of the wind respectively.  The
density profile takes on a $r^{-2}$ radial dependence when we assume a
constant wind velocity.

Very quickly, GRB fireball theorists (see Chevalier et al. 2004 for a
review) realized that this density profile, with its characteristic
$\rho \propto r^{-2}$ profile, would produce a different afterglow
light curve than, for example, a constant density profile expected
from alternate GRB models (e.g. the merger of a double neutron star
system).  With all of the evidence pushing toward massive star
progenitors for long-duration GRBs, we would expect to find that GRB
afterglows were better fit by $\rho \propto r^{-2}$ density profiles
than by constant density profiles, but roughly half of all systems
seem to be better fit by constant density profiles, with typical
densities between
0.1 and 100\,cm$^{-3}$\citep{Mes98,Dai98,Che99,PK02,Che04}.  The
disagreement between our picture of massive star progenitors for GRBs
and observations of the GRB afterglow remains an outstanding problem
in our understanding of long-duration GRBs.  In this paper, we will
study predictions of the environments surrounding
long-duration GRB progenitors in an effort to resolve the
density-profile problem.

A number of different approaches might resolve this particular
discrepancy.  On one extreme, one can ignore the observations, hoping
that future data will show a different trend and negate the current
results\footnote{A number of assumptions go into the afterglow models
needed to estimate the density profiles and it could well be that
these assumptions are incorrect.  Although roughly half of all
explosions are better fit by a wind density profile ($\rho \propto
r^{-2}$), constant density profiles can not be ruled out for any
gamma-ray burst.  But for Only 1 or 2 cases can a $r^{-2}$ density
profile be ruled out emphatically (Panaitescu - private
communication).  This observation is far from certain.}.  On the
other extreme, one can rule out the current paradigm for long-duration
gamma-ray bursts and search for a different, non-stellar progenitor.
In between a number of options exist and we will pursue two of these:
1) our free-streaming $\rho \propto r^{-2}$ profile estimate for the
density in the wind-blown region is an oversimplification; or 2) the
progenitor does not reside in its wind-blown bubble when the GRB is
produced.  We further restrict our study to progenitors of the
collapsar or collapsar-like GRB engine \citep{Woo93,Fry98,Mac99}.
This engine, which invokes the collapse of a massive star down to a
black hole and the formation of an accretion disk around that black
hole, only works for a small subset of massive star progenitors.  We
believe that by studying this small subset of progenitors, we may gain
some insight into the environments surrounding GRB engines.

The first option, modifying the density profile in the wind-blown
bubble, can be accomplished in a number of different ways and has
already been studied at some level.  Some observations can be made to
fit a wind profile if the density is decreased \citep{Wu03,Dai03}.
Chevalier et al. (2004) have also, in a generic sense, studied the
true termination radius of the Wolf-Rayet wind.  If the wind
terminates at a small radius, the observations are probing the
constant density region beyond the termination shock, and not the
$\rho \propto r^{-2}$ free-streaming wind region.  In \S 2 of this
paper, we study the wind profiles of massive stars designed to produce
collapsars (``massive star'' progenitor to the collapsar engine),
using a combination of stellar evolution models and 1-dimensional and
3-dimensional wind models.  The wind mass loss is time-dependent, and
we study the effects of this time dependence.  Most working collapsar
models require binary interactions, and we also study these
binary effects.

Alternatively, there is a collapsar progenitor scenario where, before the GRB
explosion, the progenitor moves out of its wind nebula: this is the
helium-merger progenitor (Fryer \& Woosley 1998; Zhang \& Fryer 2001).
In this progenitor, the more massive star in a binary collapses to
form a neutron star or black hole, providing a kick to its binary
system.  Not until its companion evolves off the main sequence and
causes the stars to merge is a burst produced.  Although Fryer \&
Woosley (1998) already predicted that such a progenitor would more
likely yield a surrounding environment akin to a constant density
interstellar medium (ISM) environment, no work to date has studied this
claim in detail.  In \S 3, we show results of Monte Carlo calculations
to study the positions of these bursts relative to the formation site
of their binary progenitor.  

We ultimately find that the seeming discrepancy between simple wind
models and GRB afterglow observations can play an important role in
homing in on the true progenitor of these long-duration bursts.
Although we can not rule out the massive star progenitor GRB scenario,
it is becoming increasingly difficult to match such a progenitor to
observations.  The helium merger scenario is best able to fit current
afterglow lightcurves, but it may have problems matching other
observational constraints.  We conclude by discussing the issues
remaining for the different long-duration GRB scenarios.

\section{Stellar Winds from Collapsar Progenitors}

Weaver et al. (1977) first discussed the structure of wind-driven
bubbles produced by massive stars.  Assuming that the wind profile
varies with at most a linear dependence on time and is spherically
symmetric, the structure of these wind-blown bubbles is well known
(Koo \& McKee 1992a,1992b).  However, as we shall see in this section,
neither of these assumptions is strictly true.  We study the bubble
structure produced by the rapidly varying winds from Wolf-Rayet stars,
the likely progenitor star for collapsar GRBs.

\subsection{Understanding Wind Profiles}
\label{section:parameter}

Before we study realistic wind profiles based on stellar mass-loss
histories, we first build our intuition of bubbles from winds with
constant mass-loss rates blowing into environments with different 
densities.  Such wind blown bubbles have been studied in detail
analytically (Koo \& McKee 1992a,1992b; see Dwarkadas 2005 for a
review) and the wind environment can be described by 4 separate
regions: (1) the freely-flowing fast wind, (2) the shocked fast wind,
(3) the shocked interstellar medium and (4) the unshocked interstellar
medium.  For the purpose of comparing to the afterglow observations,
the primary constraints come from the first region, where the free-flowing
winds develop a $\rho \propto r^{-2}$ density profile.  To overcome the
observational constraints, we must either lower the density of this
region or limit its size.  The free-flowing region is capped by a
termination shock ($r_{\rm termination}$), and we will focus on
determining what conditions can drive down this termination shock to
low radii: $r_{\rm termination}<0.1$\,pc.  At such small size scales,
the GRB jet can pass through this region without seriously impacting
the afterglow emission.

To understand the dependence of the termination shock radius on the
uncertainties in wind parameters, we have modified the 1-D Lagrangian
code developed by Fryer et al. (1999) to include the addition of mass
from winds.  Wind material is added in zones of constant mass.  For
constant mass loss from winds, this means that zones are added at
constant time intervals.  The inner boundary of the simulation moves
out with the wind velocity.  When a particle is added, its outer
radius is set to the inner boundary radius and the inner boundary is
set to the initial inner boundary radius.  By varying the time
intervals according to individual wind velocities, this technique can
be used to simulate winds with varying mass loss rates and varying
velocities.  We model a region from $\sim$0.01 to 20-200\,pc.  For our
1-dimensional, constant mass-loss studies, the interstellar medium is
modeled with between $\sim$50,000 and 100,000 zones.  In 250,000\,y,
roughly 25,000 zones are added to the simulation.  Table 1 summarizes
all of the parameter study models.

Figure 1 (top) shows the density profiles for four different wind
bubble simulations differing only in the density of the ambient medium
(roughly 10, 100, 1000, and 10,000 cm$^{-3}$).  We have assumed a
constant wind velocity (1000\,km\,s$^{-1}$) and constant mass loss
rate ($10^{-5}M_\odot\,y^{-1}$) which roughly corresponds to what we
would expect from a Wolf-Rayet wind.  A lot can be learned from these
simulations, but we will focus on the radius of the termination shock
($r_{\rm termination}$).  250,000\,y of this strong wind produces a
termination shock that is at a radius of 1\,pc if the wind is blowing
into an interstellar medium with a density of 100\,cm$^{-3}$.  Fitting
to these simulations, we find that the radius of this termination
shock is $\approx 1 (n_{\rm ISM}/100 {\rm cm^{-3}})^{-1/2} {\rm pc}$.
It would require very high densities indeed to place the termination
shock below 0.1\,pc.  We include a final simulation (dotted line) that
includes a cooling routine outlined in Fryer et al. (2006) based on
the cooling rates of Sutherland \& Dopita (1993).  The primary effect
of cooling is to alter the density profile near the shock with the
interstellar medium.  Another effect is the fact that the UV photons
from the star will drive an ionizing front ahead of the wind.  This
has been recently studied by van Marle et al. (2004).  Although much
more must be done to understand its effects, van Marle et al. (2004)
found that the primary change from the ionization front was 
in the outer radius of the wind shock.  It does not significantly
alter the radius of the free-streaming region and we will not
consider it further in this paper.

Figure 1 (bottom) shows the dependence of the termination shock radius
on the other free parameters in a constant wind: the mass-loss rate
(solid lines) and the wind velocity (dotted lines).  Here we have
assumed a low density interstellar medium (10\,cm$^{-3}$).  The dark
solid line assumes a mass-loss rate of $10^{-4}M_\odot\,y^{-1}$
whereas the light solid line assumes a mass-loss rate of
$10^{-6}M_\odot\,y^{-1}$.  The termination shock radius is also
proportional to the mass loss to the 1/3 power.  The dark and light
dotted lines assume wind velocities of 500\,km\,s$^{-1}$ and
2000\,km\,s$^{-1}$ respectively.  Although this changes the position
of the outer shock of the wind-blown bubble, it does not significantly
affect the radius of the termination shock.

To restrict the termination shock to a radius below 0.1\,pc, either
the ambient density surrounding the GRB progenitor must be high
($n_{\rm ISM}>10,000 {\rm cm^{-3}}$) or the Wolf-Rayet wind must have
a mass-loss rate much lower than the canonical
$10^{-5}M_\odot\,y^{-1}$.  Such constraints may not be that extreme.
Densities in excess of $10,000 {\rm cm^{-3}}$ are common in molecular
clouds, especially near massive star birthsites.  For example, the
shock producing the young supernova remnant Cas A is impacting a
650\,M$_\odot$ molecular cloud with densities of $10^7 {\rm cm^{-3}}$.
Unfortunately, such a dense medium is ruled out already by afterglow
observations (Berger et al. 2003; Chevalier \& Li 2000; Panaitescu \&
Kumar 2002; Kobayashi \& Zhang 2003)\footnote{Although bear in mind
that if we are observing the GRB shock as it progresses through
shocked wind and ISM material, the observed density will be much less
than the density of the interstellar medium.  Thus, it is possible 
that large interstellar medium densities can exit.}.

However, the Wolf-Rayet mass-loss rate could be much lower than
the canonical value for the specific stars that make GRB progenitors.
Such a low mass-loss rate might result from a helium star formed in a
binary interaction that causes a 15-20\,M$_\odot$ star to become a
Wolf-Rayet star.  Such a star can have a Wolf-Rayet mass loss rate as
low as $10^{-7}\,M_\odot\,y^{-1}$.  This is not entirely satisfactory,
as we believe such a star is not likely to collapse to a black hole.
Alternatively, because the Wolf-Rayet mass-loss rate depends upon the
metallicity ($\dot{M}_{\rm Wolf Rayet} \propto Z^{0.5}-Z^{0.9}$), GRB
progenitors with observed surroundings matching constant density
profiles could be limited to low metallicity stars (Yoon \& Langer
2005).  A low metallicity star turned Wolf-Rayet via a binary common
envelope phase could have a very low Wolf-Rayet mass loss rate.  For
example, if $\dot{M}_{\rm Wolf-Rayet} \propto (Z)^{0.9}$ (where $Z$ is
the metallicity), then a 25\,M$_\odot$ star with a metallicity 1/100th
that of solar could have mass loss rates below
$10^{-7}M_\odot\,y^{-1}$.  Not only will the free-streaming region of
this wind be small, but its total density will also be low (a better fit
to the observations).  With such stars, ambient densities of a few
hundred ${\rm cm^{-3}}$ would be sufficient to restrict the
termination shock to less than 0.1\,pc.

Before we move onto the complexities arising from realistic wind
models, we mention a few other effects that may also limit the
free-streaming wind profile in GRB progenitors.  These stars may only
form in very dense clusters and, if so, the interaction of the winds
could limit the extent of the free-streaming profile.  Rockefeller et
al. (2005) studied the hydrodynamics of winds in the Arches and
Quintuplet clusters.  By analyzing their data, we find that the
free-streaming wind region of even the strongest winds can be capped
at 0.5\,pc at the center of the cluster.  

In addition, these massive stars may have large velocities with
respect to their circumstellar environments.  The ram pressure of the
wind environment against this circumstellar environment limits the
extent of the wind.  This extent can easily be derived by setting the
free-streaming wind pressure $\equiv 0.5 \rho_{\rm wind} v_{\rm
wind}^2 = \dot{M}_{\rm wind}/(8 \pi) r_{\rm wind}^{-2} v_{\rm wind}$ 
to the ram pressure of the motion through the circumstellar medium
$\equiv 0.5 \rho_{\rm ISM} v_{\rm star}^2$.  Here $\rho_{\rm
wind},v_{\rm wind}, r_{\rm wind}$ and $\dot{M}_{\rm wind}$ are,
respectively, the density, velocity and radial extent and mass-loss
rate of the free-streaming wind; $\rho_{\rm ISM}$ is the density 
of the interstellar medium, and $v_{\rm star}$ is the motion of 
the star with respect to this circumstellar medium.  Setting 
these two ram pressures equal, we derive an equation for the 
radial extent of the wind:
\begin{equation}
r_{\rm wind} = 0.56 \left( \frac{\dot{M}_{\rm wind}}{10^{-5} {\rm M_\odot
y}^{-1}} \right)^{0.5} \left( \frac{v_{\rm wind}}{1000\,{\rm km s}^{-1}} \right)^{0.5} 
\left( \frac{\rho_{\rm ISM}}{10 {\rm cm}^{-3}} \right)^{-0.5} \left( \frac{v_{\rm star}}
{100\,{\rm km s^{-1}}} \right)^{-1}
\, {\rm pc}.
\end{equation}
The number of massive stars with velocities in excess of $100\, {\rm
km s^{-1}}$ is very low ($10-30$\%: Hoogerwerf et al. 2001) and unless
something in the progenitor evolution requires these fast velocities,
it is unlikely that these motions can explain the environments.

\subsection{Real Stellar Mass Loss}

With the intuition gained from our constant mass-loss winds, we are
set to attack the more realistic problem of stellar winds from GRB
progenitors.  In this paper, we use the output from the Tycho stellar
evolution code (Young \& Arnett 2005) and we study three different
possible GRB progenitors: a 40\,M$_\odot$ single star, a 23\,M$_\odot$
binary system and a 16\,M$_\odot$ binary system.  In the binary
systems, we assume that a common envelope phase removes the hydrogen
envelope of the primary when the primary star expands off the
main-sequence.  The mass loss rate and wind velocities for the GRB
progenitors are calculated at each timestep during the
evolution. Three prescriptions are used during steady mass loss
phases. Mass loss for blue, hydrogen-rich stars ($log T_{eff} >
3.875$) uses \citet{kud89}. At lower effective temperatures
appropriate for the red giant/supergiant phases, the prescription of
\citet{bl95} is used. For Wolf-Rayet mass-loss rates, we use the
\citet{ln03} prescription.

Catastrophic mass loss episodes are dealt with in a more schematic
fashion. The 40 $M_{\odot}$ star is subject to luminous blue variable
eruptions. The conditions for the instability (large radiative
accelerations at temperatures coinciding with ``bumps'' in the
continuum opacity) are idenitifiable in the stellar model. When these
conditions are satisfied a large amount of mass is removed from the
model to mimic an eruption. Limitations on the numerical stability of
the code prevent the mass from being removed as quickly as in a real
luminous blue variable (LBV) eruption. Instead mass is lost at a high
but steady rate dictated by code stability for a period of order 100
years with exact duration dictated by the amount of mass loss
necessary. Such episodes can be identified in Figure 2.  We choose a
conservative approach, removing only enough mass to stabilize the
stellar structure, though it is possible that the envelope could be
removed to the depth of the instability generating region. Depending
on the stellar structure, this could represent several solar masses of
ejection. In the first, large eruption $\sim 1M_{\odot}$ is
removed. Subsequent eruptions are smaller, and the bulk of the mass
loss occurs via steady winds.

The 16 and 20 $M_{\odot}$ models are intended to represent stars which
lose their hydrogen envelopes during the first ascent Red Giant Branch
(RGB) to a binary interaction. As with catastrophic eruptions, the
physical processes cannot be captured realistically in a 1D stellar
evolution code, so we must use an approximation. A similar procedure
is used, which results in an ejection of the hydrogen-rich envelope
($\sim 10 \, M_\odot$ , depending on stellar mass and mass loss history) in
less than a thousand years. From an evolutionary standpoint, this is a
good approximation, as the nuclear timescale of the star is much
greater than a thousand years at this stage. This is not necessarily a
good approximation to the mass loss, but it does provide the correct
qualitative behavior: a short period of very high density, low
velocity ($\sim v_{escape}$) mass loss which clears a cavity around
the star into which a fast Wolf-Rayet type wind expands. The star
subsequently evolves as a WR towards a Type Ib supernova. Though the
binary ejection of the hydrogen envelope is handled only in a very
schematic way it does represent, along with the subsequent evolution,
a significant improvement over constant density or power law density
profiles for the circumstellar medium.

The Wolf-Rayet mass loss is characterized by a high mass loss rate,
high velocities, and an episodic history. The abrupt changes in mass
loss occur when the main driver for the mass loss switches between
opacity bumps at different temperatures and hence different radii in
the star (see Lamers \& Nugis 2003 for a discussion of the Wolf-Rayet mass
loss mechanism). Early transient spikes may represent eruptions driven
by high radiative accelerations in the stellar atmosphere akin to LBV
or Ofpe \citep{mwd00} eruptions.

\subsection{Wind-Blown Bubbles From Real Winds}

The winds from our Tycho models have a number of characteristics that
will complicate the wind profile.  Note that the ejection velocity is
often low when the mass loss peaks (either from a LBV eruption or a
common envelop mass ejection).  The high-velocity wind ejecta that follows
these peaks will bounce against the slow-moving, dense, peak ejecta,
causing a shock to flow back down through the wind.  Also, a similar
shock will be produced when the eruption hits the termination shock.
With realistic winds, our free-streaming, ``unshocked'' wind region
now has many shocks running through it.  This will dramatically affect
the entropy profile of this wind.  But afterglow observations are not 
sensitive to the entropy, but rather the the density profile of 
the wind-blown bubble region.

To study the density profile, we first return to our 1-dimensional
simulations.  By allowing non-constant time intervals for the addition
of zones, we can incorporate the variable mass-loss rate and
time-dependent wind velocity profiles from our Tycho models.  We
assume an ambient interstellar density of $100 {\rm cm^{-3}}$.  Table
1 lists the simulations of these realistic mass-loss profiles; we give
rough ranges for the mass loss rates and velocities instead of
absolute values.  We model the wind-driven bubble only during the last
few hundred thousand years of the star's life: 0.6, 0.5, 0.6\,Myr for
the 16,23,40\,M$_\odot$ stars respectively.  Although this reduced
time does not accurately model the outer extent of the winds, it does
get the position of the free-streaming region fairly
accurately\footnote{Vikram Dwarkadas is currently modeling the full
wind profiles of our binary star in an effort to understand the
supernova remnant Cassiopeia A.  The extent of the free-streaming
region from his shock agrees with ours at the 20\% level.}.  Most of
the explosive phases of mass loss occurs during this time, and it is
this ejecta that dominates the density profile of the free-streaming
wind region.  Figure 3 shows the density profiles of these models
flowing in an ambient density of 100\,cm$^{-3}$.  For these models, it
is difficult to pinpoint the termination shock of the free-streaming
wind and there is no strict $\rho \propto r^{-2}$ dependence to the
density.  But the density does decrease roughly as $r^{-2}$ well
beyond 1\,pc.  Recall that the termination shock from our
$10^{-5}\,M_\odot\,y^{-1}, n_{\rm ISM}=100\,{\rm cm^{-3}}$ calculation
was at 1\,pc (even for the 40\,M$_\odot$ star where the mass loss
exceeds this standard value).  The ambient medium would have to exceed
$10^4\,{\rm cm^{-3}}$ to reduce this region below 0.1\,pc.  This
density is probably ruled out by the timescale of the radio emission
(see Section~\ref{section:parameter}).

We find that episodic spikes in the mass loss produce a termination
shock which causes a significant departure from an $\rho \propto
r^{-2}$ density profile. This does not help the situation much because
the region of departure from a free-streaming profile propagates out
from the star at roughly the velocity of the slow moving ejecta ($\sim
{\rm a \; few \, } 100 {\rm km\ s^{-1}}$). Thus the density profile is
disturbed within the $0.1$\,pc limit for 1000 years or less, roughly 1\%
of the time between eruptions. Although we cannot rule out such epsiodes within
1000 years of collapse unless the stellar surface is hotter than $\sim
5\times 10^5\ K$ (hotter than the continuum opacity bumps which drive
the high mass loss rates), the odds of an ejection event occurring
within 1000 years of collapse are too small to explain the
distribution of afterglows unless the frequency of ejection events
increases as the star approaches collapse. We do not observe this
behavior, but the uncertainties in the physics of mass loss do not
allow us to eliminate the possibility.

Another wrinkle in this model is the fact that these winds are far
from spherically symmetric.  As argued by Fryer et al. (1998), the
idea that GRB progenitors most likely arise from binary systems is
gaining increasing acceptance (Petrovic et al.  2005, Fryer \& Heger
2005).  If the wind is not already asymmetric due to rotation (Maeder
\& Meynet 2000), the binary interaction will ensure that it is.  If
the mass ejection arises from common envelope evolution, then this
mass outflow at least is highly asymmetric.  LBV eruptions are also
observed to be strongly bipolar.

We have modified the three-dimensional SNSPH code (Fryer, Rockefeller \&
Warren 2005) to model these winds and then study the effects of
asymmetries on the density structure.  We add particles in shells centered
on the location of the wind source, after a time $\Delta t \simeq
h/v_{wind}$ has passed since the last addition of particles, where $h$ is
the SPH particle smoothing length and $v_{wind}$ is the current value of
the wind velocity.  The mass of each particle is set according to $m =
\dot{M}_{wind} \Delta t/n_{shell}$, where $\dot{M}_{wind}$ is the current
value of the wind mass loss rate and $n_{shell}$ is the number of
particles per shell.  We remove particles from the simulation when they
pass beyond 3~pc from the wind source.

Figure 4 shows the density profile of one such SPH simulation.  A
$23\,M_\odot$ binary system, located at the center of the simulated
volume, ejects material during a common envelope phase at a rate of $\sim
0.1 M_\odot$~yr$^{-1}$.  We assume that this material is not ejected
isotropically; instead, we place it in a 45$^{\circ}$ wedge oriented
perpendicular to the $z$ axis.  The subsequent wind from the binary system
has a much higher velocity but represents a much lower rate of mass loss
from the system, so the wind collides with the inner edge of the dense
wedge and flows out of the system primarily along the $z$ axis.  In this
calculation, each added shell of wind material contains 400 SPH particles,
each shell is added at a radius of $0.01$~pc from the wind source, and the
inner edge of the dense wedge is at $r = 0.3$~pc.  The total number of
particles in the simulated volume is $\sim 350,000$ but fluctuates
slightly as the wind velocity changes.

Despite the modifications we have introduced to the wind profiles, we
have not produced any conditions that cause the density profile of the
wind-blown bubble to differ significantly from the $\rho \propto
r^{-2}$ dependence.  With these results in mind, our proposed solution
for a the massive star progenitor for GRBs is a low-metallicity,
20-25\,M$_\odot$ star turned into a Wolf-Rayet star through a binary
common envelope phase.  A 23\, M$_\odot$ progenitor with a metallicity
1/100th that of solar would have a weak Wolf-Rayet wind, but also be
massive enough to collapse to form a black hole.  A GRB in a high
metallicity host would rule out such a scenario, forcing us to
progenitor models like the He-merger scenario.

\section{He-Merger GRB Progenitors}

Before we study the environments surrounding He-merger GRB
progenitors, we first review the He-merger scenario itself.  The
He-merger progenitor for GRBs invokes a binary system where the most
massive star collapses to either a neutron star or a black hole.  If
this compact object merges with its companion, it will spiral into the
center of the companion while spinning up the companion.  In this merger, 
the binary has satisfied the three main requirements behind the collapsar 
model:  1) the compact object accretes enough to collapse to a black 
hole, 2) the star surrounding the black hole has enough angular 
momentum (perhaps too much as we will discuss below) to form a disk 
around that black hole, and 3) the hydrogen envelope is ejected in
a common envelope excretion disk.

There are two main binary evolution scenarios that dominate the
production of He-Merger progenitors of GRBs: common envelope mergers
and kick-induced mergers of nearly equal mass stars.  Figure 5
shows the basic evolutionary phases in the common envelope scenario.
In this scenario, we generally begin with an initial binary in a
fairly close orbit ($\lesssim$ 1 A.U.).  The more massive star
(``primary'') evolves off the main sequence and, for most progenitors,
envelops its companion as it turns into a giant (or supergiant) star.
This causes the orbital separation of the binary to tighten in a
common envelope phase.  The primary then collapses, forming either a
neutron star or black hole.  A fraction of these binary systems will
remain bound after the supernova explosion.  When the companion star
evolves off the main sequence, it can envelop the compact remnant
produced by the collapse of the primary.  A fraction of these binaries
will merge during this second common envelope phase, producing a
helium-merger GRB.

The second main scenario we loosely term the ``Brown'' He-merger
scenario because it uses similar initial conditions (a binary system
with nearly equal massed stars) to a double neutron star formation
scenario proposed by Bethe \& Brown (1998).  The idea of Bethe \&
Brown (1998) was to consider binary systems where the two stellar
components have masses that are so close that the less massive star
evolves off the main sequence before the more massive star collapses
to form a compact remnant (this corresponds to masses of the two stars
that are within $\sim$10\% of each other).  In their model, the two
helium stars produced during the mass transfer phase are pushed into a
tight binary that remains bound through the two subsequent supernova
explosions, forming a system with two compact objects.  In our
scenario, we follow the Bethe \& Brown (1998) evolution up to the
point that we form two helium stars in a tight binary (Fig. 6).  
When the first helium core collapses, its kick can drive it 
into its companion, producing a helium-merger GRB.

This ``Brown'' He-merger scenario uses supernova kicks to cause the
compact object to merge with its stellar companion.  This begs the
question of whether a kick imparted to the compact object formed
during the collapse of the primary can drive the neutron star or black
hole into its main sequence companion (Fig. 7).  Because main sequence
stars are not quite as compact as helium stars, we expect lower
accretion rates in these models, possibly making this scenario less
viable as a GRB progenitor.  In addition, Fryer et al. (1999) pointed
out that this main-sequence star mechanism is rare.  We will
explicitly discuss the rates here, finding that this mechanism accounts for
less than $\sim$5\% of the total helium-merger GRB population for most
simulations.

These three different progenitor scenarios produce very different
environments through which the GRB jet will pass (Table 2).  Both
mechanisms 2 and 3 produce mergers very soon after (generally within a
few years of) the supernova explosion.  GRB explosions for these two
scenarios will first pass through a young supernova remnant and then
through the wind from the Wolf-Rayet progenitor.  Scenario 1 does not
produce a GRB until long after ($\sim 1-10$\,Myr) the collapse of the
primary star.  With the kick imparted to the system during the
collapse of the binary, most bursts from this progenitor occur beyond
the edge of the supernova remnant and the Wolf-Rayet wind termination
shock.  Hence, the environment surrounding these bursts will not be
the wind-swept structure characteristics we expect from Wolf-Rayet
stars.  The immediate environment surrounding these bursts will either
be characterized by a constant density medium (if the companion is
low-mass: $< 15-20$\,M$_\odot$) or by the giant winds from the
companion.  In this section, we study the relative frequencies of
occurrence of these different burst scenarios and their velocity,
merger time, and position distributions including a brief study of the
uncertainties of all these values.

\subsection{Possible Constraints on the Helium merger}

It is likely that only a subclass of He-mergers will produce GRBs.
Currently we do not understand the progenitors or the collapsar
mechanism well enough to definitively exclude some progenitors.
Perhaps observations of GRBs can provide some insight into the
progenitor subclass that truly makes GRBs.  Here we discuss a variety
of He-merger subclasses that may make up the true set of GRB
progenitors.  We will then study each of these subclasses separately
when we present our population synthesis results.  Table 3 summarizes
these different subclasses.

One of the sources of uncertainty in our population synthesis studies
is that we don't know exactly what conditions actually produce GRBs.
For example, Fryer \& Woosley (1998) assumed that the primary could
collapse into either a neutron star or a black hole and still be a
candidate GRB progenitor.  Their assumption, which has been borne out
by current simulations (Zhang \& Fryer 2001), is that even if the
compact remnant were a neutron star, it would accrete so much in the
inspiral phase that it would become a black hole quickly, satisfying
the ``black hole requirement'' producing collapsars.  But if the
accretion rate is not so extreme (angular momentum, for example, could
drive outflows), then the neutron star may not collapse to form a
black hole and such a progenitor will not produce a collapsar.
Alternatively, it may be that black holes accrete too much mass,
making inefficient collapsars unable to drive strong GRB jets.  We
will look separately at systems whose primary collapses to a neutron
star versus those systems whose primary collapses to form a black
hole.  This gives us two separate subclasses based on the initial status
of the compact remnant: 1) a neutron star or 2) a black hole.

Another constraint on the progenitor may be the amount of angular
momentum.  When Zhang \& Fryer (2001) modeled their mergers, they
focused on maximizing the angular momentum surrounding their collapsar
black hole.  They assumed that the stars were, if anything, spinning
in the same direction as the orbital angular momentum.  They also
assumed that very little orbital angular momentum was required to
remove the hydrogen envelope and that all of the orbital angular
momentum was deposited in the helium star.  This produced very high
angular momentum disks in the merged black hole accretion disk system,
leading some to argue that this progenitor does not produce a viable
GRB \citep{DiM02}.  However, recall that Zhang \& Fryer (2001) were
trying to maximize the angular momentum in the system.  The angular
momentum could be much lower if most of the orbital angular momentum
is lost to the hydrogen envelope or if the star is spinning in the
opposite direction with respect to the binary orbit.  Mergers produced
with kicks may also lead to lower angular momenta.  In general, we
assume that a star's spin direction in a binary coincides with the
direction of the orbit.  This is almost certainly true prior to the
supernova explosion.  But the kick imparted to the compact object
during supernova explosion can (and often does for those binaries that
remain bound) produce binaries where the spin angular momentum
direction is 180 degrees off from the orbital angular momentum of the
binary.  This subclass is defined in our calculations by restricting
our sample to those companion stars whose helium star moment of
inertia (from the helium stars of Fryer \& Heger 2005, $I_{\rm
He-Star} \approx 0.1 M_{\rm He-Star} R^2_{\rm He-Star}$) is greater
than the moment of inertia of the merging compact object when it
reaches the helium star radius ($I_{\rm Compact Binary} = M_{\rm
Compact Object} R^2_{\rm He-Star}$).

Lastly, if the companion star is too massive, it too will blow a wind,
leading to a surrounding wind environment not too different from the
Wolf-Rayet winds produced by the primary star.  We define this
subclass of low-mass losing systems by restricting our sample to
systems with companions whose total mass-loss is less than
1\,M$_\odot$.  This essentially limits us to binaries where the
initial mass of the companion is less than 15\,M$_\odot$.

With these four subclasses in mind, we can now analyze our population 
synthesis studies.

\subsection{Population Synthesis studies}

Our population synthesis calculations are based on the code described
in Fryer et al. (1998,1999).  The main alteration to this code has
been to update the mass of the collapsed core using the results of
Fryer \& Kalogera (2001).  Fryer et al. (1999) set the remnant mass
for stars collapsing to neutron stars to 1.4\,M$_\odot$ and the
remnant mass for stars collapsing to black holes to one third the mass
of the star at collapse.  In this paper, we have fit a function to the
results of Fryer \& Kalogera (2001), allowing a more continuous spread
of neutron star and black hole masses.  We refer the reader to Fryer
et al. (1998,1999) for any other details about the population
synthesis technique or the specific code we use.  We have run over 
a dozen models using a wide range of choices for the uncertain 
binary population synthesis parameters (Table 4).

Table 4 shows a list of all the population synthesis models studied in
this paper followed by the values of the population synthesis
parameters used by each model.  For a more detailed description of
each of these parameters, see Fryer et al. (1999).  We focus on just a
few parameters: the power law of the initial mass function
($\alpha_{\rm IMF}$), the distribution of velocities imparted to
compact objects at collapse (kick), mass transfer parameters denoting
the fraction of overflowing matter in a binary accreted by the
companion ($\beta_{\rm MT}$) and the specific angular momentum of the
overflowing matter ejected, and hence lost, from the binary
($\alpha_{\rm MT}$), and the efficiency with which the inspiralling star
in a common envelope phase ejects the envelope ($\alpha_{\rm CE}$).
We sample both our primary and secondary stars from the same initial
mass function, constraining the secondary to be less massive than the
primary.  After some argument about the distribution of kicks arising
from stellar collapse, it is now generally accepted that a bimodal kick
distribution like that proposed by Fryer, Burrows, \& Benz (1998) is
roughly correct \citep{Arz02}.  We term the kick distribution
represented by two Maxwellians of 50 and 500\,km\,s$^{-1}$ the FBB
kick distribution.  We have also used a set of single Maxwellian
distributions covering a range of mean values to better understand the 
of the dependence on the kick magnitude.  Finally, we have run a few
tests of the dependencies on the stellar mass loss rates and radii.
We model 50 million binaries in each simulation.

Fryer et al. (1999) presented very few results outlining the
characteristics of the Helium Merger model, focusing mostly on the
formation rate.  Here we study the distribution of merger velocities,
merger times and distances traveled before GRB outburst for the
various subclasses of He-merger progenitor.  The explosion and the
supernova kick impart a velocity to the binary (if it remains bound).
Figure 8a shows the distribution of the post-collapse velocities of
He-Mergers for our ``Standard'' model (Table 4).  The dark solid line
shows the number of binaries per 1\,km\,s$^{-1}$ bin for all He-merger
progenitors.  Although for most subclasses, the bulk of the systems
have velocities below 25\,km\,s$^{-1}$ (where the total number peaks),
a second peak exists above 100\,km\,s$^{-1}$.  The two peaks arise
because of the double peaked nature of the supernova kick
distribution.  The peak at low velocities is not present for two
subclasses of He-merger models: those with initially black hole
compact remnants and those with high moments of inertia.  Note that
the peak in low-velocity systems is dominated by the NS Collapse and
Low-Wind models.  Because the kick must nearly exactly counter the
momentum lost in the supernova ejecta for our strong supernova models,
the systems that are more likely to remain bound for these NS binaries
are those where the countering effects lead to small proper motions.
None of the kick velocities are high enough to escape their host
galaxy's gravitational potential unless the galaxy is extremely small.

Figure 8b shows the merger times for the same model as Fig. 8a.  The
peaks below 100 years arises from binaries following the ``Brown
Merger'' scenario or the ``MS Merger'' scenario (Scenarios II and
III)\footnote{The peak between 10-100\,y is produced by the Brown
scenario in systems where the helium cores do not go through a common
envelope phase but are kicked in such a manner to merger - very rare
indeed}.  For most of these systems, the stars merge within a few
months of the collapse of the primary.  For Scenario I, the merger
rates are determined by the remaining main-sequence lifetime of the
companion.  Note that the low wind (small mass) subclass dominates the
long merger times, but nearly all of the subclasses have the full
range of merger times.  Figure 8c shows the distribution in distance
traveled before outburst for the He-merger population.  This
distribution is determined by multiplying the velocity of the binary
times its merger time.  Here again, the $<$1\,pc bin is made up
entirely of the ``Brown'' scenario and the main-sequence kick
scenario.

In Figures 8d-f, we show the same set of data for the velocity, merger
time, and distance distributions as a function of fractions relative
to the total number in a subclass instead of total number.  Because the
rate of these mergers is still very difficult to determine due to a
number of uncertainties (e.g. binary mass fraction and initial mass
function), such relative abundances are most appropriate for comparison to
observations.  From these plots we can see that in the total He-merger
population, less than 20\% occur within 1\,pc of their initial
formation site.  This is essentially a statement that less than 20\%
of He-merger systems are formed in the Brown and MS Merger
scenarios.  The moment of inertia subclass produces the most extended
distribution of mergers.

To demonstrate the effect of the kick distribution, we compare to a
single Maxwellian kick distribution.  Figure 9a-c shows the fractional
distributions of velocity, merger time, and distance travelled for our
Max300 mode.  The scales have been kept identical to those used in
figures 8d-f and these simulations can be compared directly.  Note the
lack of the double peak in the velocity distribution.  Figures 10 and
11 show the velocity, merger time, and distance travelled
distributions for a series of models, but focusing only on the total
He-merger population.  Figure 10 focuses on different mass transfer
and binary parameters - the two most important being stellar radii and
stellar mass loss.  Figure 11 shows the results of a series of models
using single Maxwellian kick distributions of varying strengths.

Tables 5,6 and 7 summarize the results of all of the simulations.
Table 5 shows the merger rates assuming that a massive binary is
formed every 100y.  In general, the standard He-merger scenario
dominates $>75$\% of all the He-mergers formed.  We confirm the
results of Fryer et al. (1999) that the main sequence merger rate is
very low ($< 5-10$\% of all mergers).  Given the high uncertainty in the
rate, we can not rule out any scenario or subclass, or claim any
subclass as the sole formation scenario or subclass based on the GRB
rate, but both the main sequence merger scenario and the high moment 
of inertia subclass definitely have rates on the low end to be the sole
path for GRB formation.  Table 6 lists velocity and merger time
information.  High velocity systems are rare, so any observations that
argue for a high velocity in the merging system would place strong
constraints on some of the population synthesis uncertainties.  The
strongest constraint at the moment may well be the number of systems
that merge within a wind background versus systems in a more constant
medium.  The effective observational constraint on this is the
distance these systems travel before merging (Table 7).  Systems with
stars that initially collapse to form black holes tend to move faster
but merge quicker, and are likely to be closer to their formation
sites.  As the observational implications get stronger, it is possible
that we will be able to home in on the true GRB progenitor subclass.

Some final summarizing comments on the nature of these populations:
\begin{itemize} 
\item{Roughly 20\% (15\% if we exclude main-sequence mergers as they
may not be dense enough to produce GRBs) of all He-mergers will occur
with the wind-driven shroud produced by the collapse of the
star forming the compact remnant for the GRB.  These Bursts will not
only occur within the wind, but also within a young supernova remnant
($< 10$\,y old)}.
\item{The rest of the systems produce bursts $\sim$1-50\,million 
years after the collapse of their massive component, roughly 
10-100\,million years after the initial burst of star formation.}
\item{Although some bursts can occur 1\,kpc away from their initial 
star formation, the bulk occur within 100\,pc of their formation sites.}
\end{itemize}
What do these results tell us about the environments around He-Merger
GRBs?  Although the star forming the compact object in these systems
may have had a Wolf-Rayet wind profile, because of the delay in the
explosion, the environment surrounding the system during the GRB
outburst need not have this $r^{-2}$ density profile.  If all of the
progenitors we discussed here make GRBs, 80\% of He-mergers will {\it
not} be characterized by a strong Wolf-Rayet wind environment.  If the
companion to the compact remnant exhibits strong winds, its wind
profile will dominate the surrounding medium.  But, in our
simulations, 60-90\% of the companions have masses below 15$_\odot$
and hence will not have winds that strongly affect the surrounding
medium.  In comparing to GRB afterglows, the surrounding media of
these systems (80\% $\times$ 60-90\% $=$ 50-70\% of all systems) will
exhibit constant density profiles.  But take these fractions with some 
caution - we do not know what subset of all He-mergers will actually 
produce GRBs, and depending upon this subset, the relative rates 
will change.

\section{Conclusion}

We have studied the environments around collapsar-like long-duration
gamma-ray bursts.  We have focused on two classes of progenitors that
have been proposed to explain these GRB explosions with their
associated supernovae: the collapse of a massive star and the merger of
a compact object with its stellar companion.  The progenitor class
leading to massive star collapse is characterized by progenitors whose
mass loss prior to collapse is extremely high.  The environment
surrounding these progenitors is defined by this mass loss rate.  The
general assumption (e.g. Chevalier et al. 2004) is that the density
profile follows an $r^{-2}$ profile with a density proportional to a
constant mass loss.  But the mass loss for these massive
stars is anything but constant.  This variability leads to
modifications to the $r^{-2}$ profile.  In addition, the mass loss 
can be significantly asymmetric, leading to additional wrinkles in the
profile of the surrounding environment.  In general, these
modifications are not sufficient to resolve the discrepancies between
predictions of this progenitor scenario and the observations.  
Recent work studying mass loss in GRB progenitors has indicated 
that these environment observations contain a wealth of information 
about the bursts (van Marle et al. 2005, Eldridge et al. 2006).

Although further study (both theoretical and observational) is
required, we are now pushed to some extreme predictions in order for the
classic massive star progenitor scenario for GRBs to produce roughly
half of all GRB explosions within constant density environments.  In
addition, the low densities predicted by some GRB progenitors will be
much harder to produce with massive star progenitors.  This, in
effect, is strike two for the massive star collapse progenitor for
long-duration GRBs -- the first strike being the difficulty in getting
enough angular momentum to form black hole accretion disks in these
progenitors (e.g. Petrovic et al. 2005, Fryer \& Heger 2005).  A
possible solution may still exist if GRB progenitors are low-mass
(20-25\,M$_\odot$), low-metallicity stars.  The resultant Wolf-Rayet
winds for such stars could be less than $10^{-7}M_\odot\,y^{-1}$.  

The other class of progenitor, the He-merger model, invokes the
merger of a compact remnant from a massive star with its companion.
Although this progenitor model also begins with a massive star with
strong winds, the system can travel up to 1\,kpc from this wind
environment prior to the GRB explosion.  We described 3 different
formation scenarios: the classic ``He-merger'' model produced when the
companion envelops its compact remnant companion, the Brown merger
scenario requiring nearly equal-massed binary stars, and the
main-sequence kick scenario.  The relative rates of these three
scenarios are roughly 80:15:5 respectively.  Essentially all of the
classic He-merger systems explode more than 10\,pc from their initial
formation site.  For these models, the surrounding environments are not
set by any wind profile (unless the companion star also has a strong
wind).  In contrast, the Brown and main-sequence merger scenarios
occur within the wind profile of the compact remnant forming star.
Indeed, the GRB from these scenarios will occur within 10-100\,y of
the collapse of the primary (and its possible explosion).  Assuming that
these scenarios produce all long-duration GRBs, we predict at least
20\% of all GRBs should have wind profiles.  The remaining 80\% will
exhibit weaker wind environments (not from Wolf-Rayet stars) of the
NS/BH companion, 60-90\% of which are stars with masses below
15\,M$_\odot$.  This means that, if all He-mergers produce GRBs, we
expect that 50-70\% of all GRBs should occur in ISM-like surroundings.
Roughly 20-30\% of these bursts will occur more than 100\,pc from
their initial formation sites.

It appears that the He-merger class of long-duration progenitors fits 
the constant density surroundings better than the stellar collapse 
progenitors.  This is not surprising.  Fryer \& Woosley (1998) 
proposed this model, in part, to explain this observation.  However, 
this model still has one strike against it:  these mergers tend to 
have too much angular momentum.  Note also that these progenitors 
predict significant hydrogen in an excretion disk not too far from 
the collapsing star.  Although one would  not expect hydrogen 
lines to be strong in the supernova explosion accompanying the 
GRB from this He-merger progenitor, ultimately evidence of 
the supernova shock interacting this disk may produce some 
hydrogen emission akin to the hydrogen line emission observed 
in the Type Ia supernova 1999ee (Mazzali et al. 2005).

If He-merers are the only long-duration GRB progenitor, these
constraints will probably require that only a subset of all the
possible He-merger models actually produce GRBs, but much more
detailed calculations (both of population synthesis and of common
envelope mergers) are required to constrain this subset.  Observations
combined with population synthesis studies may provide clues to the
true subclass of He-mergers responsible for the formation of GRBs.

If observations can constrain the subclass of He-mergers that produce
GRBs, we may also be able to constrain some of the uncertainties
currently plaguing population synthesis calculations.  We can use
these comparisons to help validate our population synthesis codes for
use in other binary calculations: e.g. the prediction of populations
of X-ray binaries.  These calculations also ultimately teach us about
stellar evolution.  The most clear-cut examples of uncertain
parameters in this study are stellar radii and stellar winds.  In this
way, studying the environments around GRBs not only teach us about
GRBs, but also the objects that make GRBs.

Alternatively, both progenitor scenarios may produce GRBS:  the He-merger 
model would be responsible for producing all $\rho=$constant ambient 
medium bursts along with some of the $\rho \propto r^{-2}$ bursts; 
the classic massive star progenitor will account for a fraction of 
the $\rho \propto r^{-2}$ wind-blown bubble bursts.

{\bf Acknowledgments} This work was funded in part under the auspices
of the U.S.\ Dept.\ of Energy, and supported by its contract
W-7405-ENG-36 to Los Alamos National Laboratory, by a DOE SciDAC grant
DE-FC02-01ER41176 and NASA Grant N5-SWIFT05-0014.  

{}

\newpage

\begin{deluxetable}{lcccccc}
\tablewidth{0pt}
\tablecaption{Wind Parameter Study\label{table:windparam}}
\tablehead{
  \colhead{Simulation}
&  \colhead{$N_{\rm zone}^{0y} (N_{\rm zone}^{0.25Myr})$}
&  \colhead{$\dot{M}$}
&  \colhead{$n_{\rm ISM}$}
&  \colhead{$v_{\rm wind}$}
&  \colhead{$r_{\rm termination}$}
&  \colhead{$r_{\rm outer}$} \\
  \colhead{Name}
&  \colhead{($10^{4}$ zones)}
&  \colhead{($M_\odot y^{-1}$)}
&  \colhead{(cm$^{-3}$)}
&  \colhead{($1000 {\rm km s^{-1}}$)}
&  \colhead{($r_{\rm termination}$)}
&  \colhead{($r_{\rm outer}$)} 
}

\startdata

$\rho 10$ & 6.2(8.7) & $10^{-5}$ & 10 & 1 & 2.4 & 12.0 \\
$\rho 100$ & 7.5(10.0) & $10^{-5}$ & $10^2$ & 1 & 1.0 & 7.7 \\
$\rho 100$cool\tablenotemark{a} & 7.5(10.0) & $10^{-5}$ & $10^2$ & 1 & 1.0 & 7.7 \\
$\rho 1000$ & 9.3(11.8) & $10^{-5}$ & $10^3$ & 1 & 0.42 & 5.1 \\
$\rho 10000$ & 12.1(14.6) & $10^{-5}$ & $1.8 \times 10^4$ & 1 & 0.12 & 3.2 \\
$\dot{M}\times10$ & 6.2(8.7) & $10^{-4}$ & 10 & 1 & 4.8 & 18.4 \\
$\dot{M}/10$ & 6.2(8.7) & $10^{-6}$ & 10 & 1 & 1.1 & 7.4 \\
$v_{\rm wind} \times 2$ & 6.2(8.7) & $10^{-5}$ & 10 & 2 & 2.0 & 16.0 \\
$v_{\rm wind}/2$ & 6.2(8.7) & $10^{-5}$ & 10 & 2 & 2.0 & 13.1 \\
$16\,M_\odot$ & 2.3(2.4) & $10^{-7}-10^{-5}$ & 100 & 2-3 & $\sim 45$\tablenotemark{b} & N/A \\
$23\,M_\odot$ & 2.3(5.0) & $10^{-7}-10^{-1}$ & 100 & 0.2-3.5 & $\sim 2$\tablenotemark{b} & N/A \\
$40\,M_\odot$ & 2.3(2.8) & $10^{-6}-10^{-1}$ & 100 & 0.3-4.0 & $\sim 2$\tablenotemark{b} & N/A \\

\enddata

\tablenotetext{a}{This simulation uses the cooling routine outlined 
in Fryer et al. (2006) based on the cooling rates of Sutherland 
\& Dopita (1993).}
\tablenotetext{b}{The termination shock is less well defined for these
models.  Instead we give the radius at which a roughly $\rho \propto r^{-2}$
density profile ends.  There is certainly structure within this
radius, but probably not a sufficient deviation to change the
conclusions of the afterglow constraints.}

\end{deluxetable}

\newpage

\begin{deluxetable}{lcccc}
\tablewidth{0pt}
\tablecaption{Merger Scenarios\label{table:scen}}
\tablehead{
  \colhead{Name}
&  \colhead{Merger Mechanism\tablenotemark{a}}
&  \colhead{Companion\tablenotemark{b}}
&  \colhead{Merger Time\tablenotemark{c}}
&  \colhead{Distance\tablenotemark{d}}
}

\startdata

He-Merger\tablenotemark{e} & CE & Evolved & $10^6-10^7$\,y & $\sim 10 - \sim 1000$\,pc \\
Brown Merger\tablenotemark{f} & Kick & Evolved & $<100$\,y & $<0.1$\,pc \\
MS Merger\tablenotemark{f} & Kick & Main-Sequence & $<100$\,y & $<0.1$\,pc \\

\enddata

\tablenotetext{a}{The NS/BH merges with its companion either 
by being kicked into it during the supernova explosion (Kick) or 
by merging when the secondary expands off the main-sequence surrounding 
the NS/BH in a common envelope (CE).}
\tablenotetext{b}{We differentiate the companion based on whether it has 
an evolved helium core or the unevolved core of a main-sequence 
star.  The density in the evolved helium core can be much higher 
than that of a main-sequence star.}
\tablenotetext{c}{Time between the collapse of the primary (forming 
the compact remnant) and the merger of this compact remnant with 
the companion star.}
\tablenotetext{d}{Distance traveled by the system after the primary 
collapse.}
\tablenotetext{e}{\cite{Fry98}}
\tablenotetext{f}{\cite{Fry99}}

\end{deluxetable}

\begin{deluxetable}{lcc}
\tablewidth{0pt}
\tablecaption{Subclasses\label{table:subclass}}
\tablehead{
  \colhead{Name}
&  \colhead{Constraint}
&  \colhead{Trend in Progenitors}
}

\startdata

NS Collapse & $M_{\rm Primary} > 20 M_\odot$ & Low-Mass Primary \\
BH Collapse & $M_{\rm Primary} < 20 M_\odot$ & High-Mass Primary \\
Low-Wind & $M_{\rm Secondary} < 15 M_\odot$ & Low-Mass Secondary \\
Angular Momentum & $I_{\rm He Star} > I_{\rm Compact Binary}$ & High-Mass Secondary \\

\enddata

\end{deluxetable}

\newpage

\begin{deluxetable}{lccccc}
\tablewidth{0pt}
\tablecaption{Simulation Properties\label{table:sims}}
\tablehead{
  \colhead{Simulation}
&  \colhead{$\alpha_{\rm IMF}$\tablenotemark{a}}
&  \colhead{$v_{\rm kick}$\tablenotemark{b}}
&  \colhead{$\alpha_{\rm MT},\beta_{\rm MT}$\tablenotemark{c}}
&  \colhead{$\alpha_{\rm CE}$\tablenotemark{d}} 
&  \colhead{Other} 
}
\startdata

Standard & 2.35 & FBB & 1.0,0.8 & 0.5  & \\
StanCE0.5  & 2.35 & FBB & 0.5,0.5 & 0.5 & \\
StanCE1.0   & 2.35 & FBB & 0.5,0.5 & 1.0 & \\
StanCE0.2   & 2.35 & FBB & 0.5,0.5 & 0.2 & \\
IMF2.7  & 2.7 & FBB & 1.0,0.8 & 0.5  & \\
IMF2.7CE0.5 & 2.7 & FBB & 0.5,0.5 & 0.5 & \\
IMF2.7CE1.0 & 2.7 & FBB & 0.5,0.5 & 1.0 & \\
IMF2.7CE0.2 & 2.7 & FBB & 0.5,0.5 & 0.2 & \\
IMF2.7MT1.0 & 2.7 & FBB & NA,1.0 & 0.5 & \\
FBBHighWind & 2.7 & FBB & NA,1.0 & 0.5 & $\dot{M}_{\rm Wind} = 10 \times \dot{M}_{\rm Wind}^{\rm Stan}$ \\
FBBLowRad & 2.7 & FBB & NA,1.0 & 0.5 & $R_{\rm Star} = 0.1 \times R_{\rm Star}^{\rm Stan}$ \\
FBBHighRad & 2.7 & FBB & NA,1.0 & 0.5 & $R_{\rm Star} = 4 \times R_{\rm Star}^{\rm Stan}$ \\
Max50  & 2.7 & 50 & 0.5,0.5 & 0.5 & \\
Max100  & 2.7 & 100 & 0.5,0.5 & 0.5 & \\
Max150  & 2.7 & 150 & 0.5,0.5 & 0.5 & \\
Max200  & 2.7 & 200 & 0.5,0.5 & 0.5 & \\
Max300  & 2.7 & 300 & 0.5,0.5 & 0.5 & \\
Max400  & 2.7 & 400 & 0.5,0.5 & 0.5 & \\
Max500  & 2.7 & 500 & 0.5,0.5 & 0.5 & \\
Max300IMF  & 2.35 & 300 & 0.5,0.5 & 0.5 & \\
MaxHighWind & 2.7 & 300 & NA,1.0 & 0.5 & $\dot{M}_{\rm Wind} = 10 \times \dot{M}_{\rm Wind}^{\rm Stan}$ \\

\enddata 
\tablenotetext{a}{For our models, we sample both the primary
and the secondary from an initial mass function.  We set the minimum
primary mass to 9\,$M_\odot$.  The secondary is allowed to be as low as
1\,$M_\odot$ and we insure that its mass is less than the primary.}
\tablenotetext{b}{We use two kick distributions:  a bimodal Maxwellian
with mean velocities at 50 and 500\,km\,s$^{-1}$ based on the bimodal
distribution from Fryer, Burrows, \& Benz (1998) denoted FBB, and a
single Maxwellian denoted by the mean velocity.}
\tablenotetext{c}{Mass transfer parameters: $\beta_{\rm MT}$ is the
fraction of the expanding star's envelope that is accreted onto the
companion (the rest is assumed to be lost from the system),
$\alpha_{\rm MT}$ is the fraction of the specific angular momentum
carried away by the mass that is lost.  These parameters both
determine the mass accreted by the companion and the orbital
separation of the binary.}  
\tablenotetext{d}{If the expanding star envelops its companion, the
companion will spiral in through the envelope.  $\alpha_{\rm CE}$ is a
parameter designed to represent how efficiently the orbital energy
ejects the envelope.  It includes assumptions both about the binding
energy of the expanding envelope and the amount of orbital energy
injected during the inspiral.  If this value is low, the final
separation after a common envelope phase will also be small.}

\end{deluxetable}

\clearpage
\begin{deluxetable}{lccccccc}
\tablewidth{0pt}
\tablecaption{Merger Rates (Myr$^{-1}$)\label{table:mrate}}
\tablehead{
  \colhead{Simulation}
&  \multicolumn{3}{c}{Formation Scenario}
&  \multicolumn{4}{c}{Subclass}\\
&  \colhead{Helium} 
&  \colhead{Brown} 
&  \colhead{MS} 
&  \colhead{NS} 
&  \colhead{BH} 
&  \colhead{Low-Wind} 
&  \colhead{Ang. Mom.} 
}
\startdata

Standard & 150 & 26 & 7 & 130 & 60 & 150 & 7.8 \\
StanCE0.5  & 100 & 25 & 7.3 & 74 & 60 & 95 & 7.8 \\ 
StanCE1.0   & 99 & 23 & 7.2 & 71 & 58 & 96 & 8.6 \\ 
StanCE0.2   & 100 & 28 & 7.2 & 77 & 59 & 91 & 5.6 \\ 
IMF2.7  & 160 & 19 & 4.5 & 140 & 40 & 160 & 5.0 \\ 
IMF2.7CE0.5 & 103 & 19 & 4.5 & 82 & 45 & 98 & 5.8 \\ 
IMF2.7CE1.0 & 96 & 18 & 4.4 & 68 & 40 & 95 & 5.6 \\ 
IMF2.7CE0.2 & 100 & 20 & 4.3 & 84 & 40 & 93 & 3.6 \\ 
IMF2.7MT1.0 & 68 & 19 & 4.5 & 51 & 41 & 65 & 5.0 \\ 
FBBHighWind & 160 & 4.3 & 0 & 149 & 12 & 149 & 0.24 \\ 
FBBLowRad & 2.7 & 38 & 0 & 21 & 20 & 6.6 & 1.1 \\ 
FBBHighRad & 110 & 5.4 & 19 & 100 & 35 & 120 & 4.4 \\ 
Max50  & 0.13 & 0.026 & 0.0064 & 0.11 & 0.056 & 0.13 & 0.0066 \\ 
Max100  & 120 & 30 & 7.5 & 96 & 65 & 120 & 8.1 \\ 
Max150  & 72 & 22 & 6.7 & 52 & 49 & 69 & 6.3 \\ 
Max200  & 46 & 16 & 5.4 & 32 & 36 & 44 & 4.7 \\ 
Max300  & 18 & 7.9 & 3.0 & 12 & 17 & 17 & 2.4 \\ 
Max400  & 5.7 & 3.5 & 1.3 & 3.4 & 7.1 & 5.1 & 1.1 \\ 
Max500  & 1.5 & 1.4 & 0.46 & 0.78 & 2.5 & 1.3 & 0.38 \\ 
Max300IMF  & 21 & 11 & 4.9 & 11 & 26 & 29 & 3.8 \\ 
MaxHighWind & 19 & 1.4 & 0.0 & 19 & 1.3 & 18 & 0.04 \\ 

\enddata 

\end{deluxetable}

\clearpage
\begin{deluxetable}{lccccc}
\tablewidth{0pt}
\tablecaption{Velocity and Merger Times}
\tablehead{
  \colhead{Simulation}
& \multicolumn{4}{c}{Velocity $>50 {\rm km \, s^{-1}}$ ($>100 {\rm km \, s^{-1}}$)\tablenotemark{a}}
& \colhead{Mean Merger Time\tablenotemark{b}} \\ 
&  \colhead{NS} 
&  \colhead{BH} 
&  \colhead{Low-Wind} 
&  \colhead{Ang. Mom.} 
&  \colhead{($1-\sigma$ deviation)}
}
\startdata

Standard &  0.31(0.04) & 5.9(1.4) & 5.0(0.47) & 0.549(0.019) & 9.9(5.6) \\
StanCE0.5 & 0.54(0.062) & 4.1(0.53) & 4.9(0.46) & 0.83(0.030) & 10.4(7.9) \\
StanCE1.0 & 0.54(0.065) & 2.9(0.15) & 4.4(0.41) & 0.83(0.030) & 10.6(8.1) \\
StanCE0.2 & 0.48(0.062) & 7.7(1.7) & 6.8(0.68) & 0.82(0.033) & 9.9(7.3) \\
IMF2.7  & 0.26(0.030) & 5.3(1.1) & 5.0(0.41) & 0.36(0.011) & 10.4(5.5) \\
IMF2.7CE0.5 & 0.46(0.056) & 4.1(0.42) & 5.4(0.39) & 0.70(0.018) & 11.0(7.9) \\ 
IMF2.7CE1.0 & 0.49(0.061) & 2.7(0.12) & 4.3(0.34) & 0.61(0.016) & 11.3(7.9) \\
IMF2.7CE0.2 & 0.42(0.055) & 7.0(1.5) & 6.6(0.60) & 0.57(0.018) & 10.4(7.1) \\
IMF2.7MT1.0 & 0.74(0.088) & 3.0(0.13) & 4.8(0.40) & 0.87(0.024) & 10.7(8.8) \\
FBBHighWind & 0.077(0.0024) & 0(0) & 0(0) & 0(0) & 10.4(5.4) \\
FBBLowRad & 24(14) & 24(10) & 12(2.0) & 3.6(0.63) & 9.5(9.9) \\
FBBHighRad & 0.009(0) & 2.4(0.096) & 3.2(0.11) & 0.25(0.0018) & 10.1,4.8 \\
Max50  & 0(0) & 0.72(0) & 3.0(0) & 0.32(0) & 11.1,(7.0) \\
Max100  & 0.42(0.0023) & 2.3(0.0046) & 2.9(0.0074) & 0.56(0.0034) & 11.0(8.1) \\
Max150  & 1.1(0.0054) & 4.5(0.015) & 5.6(0.015) & 1.4(0.0015) & 11.0(8.6) \\ 
Max200  & 3.1(0.016) & 11(0.036) & 13(0.043) & 3.7(0.0041) & 11.0(8.8) \\
Max300  & 11(0.10) & 35(0.27) & 40(0.21) & 15(0.024) & 11.0(9.2) \\
Max400  & 23(1.2) & 49(1.7) & 55(1.1) & 24(0.19) & 10.9(9.8) \\
Max500  & 44(9.6) & 65(12) & 69(7.8) & 39(1.8) & 11.0(10.7) \\
Max300IMF & 13(0.13) & 37(0.28) & 41(0.22) & 19(0.034) & 10.1(9.3) \\ 
MaxHighWind & 2.4(0.0052) & 0.015(0) & 0.54(0) & 0(0) & 10.4(5.4) \\

\enddata 
\tablenotetext{a}{Percentage of systems in each subclass that have  
velocities greater than $>50 {\rm km \, s^{-1}}$ ($>100 {\rm km \, s^{-1}}$)}
\tablenotetext{b}{Mean merger time in Myr ($1-\sigma$ deviation in this merger 
time assuming Gaussian distribution).}

\end{deluxetable}

\clearpage
\begin{deluxetable}{lcccc}
\tablewidth{0pt}
\tablecaption{Distances Travelled between collapse and Merger}
\tablehead{
  \colhead{Simulation}
& \multicolumn{4}{c}{Distances $>10$\,pc ($>100$\,pc)\tablenotemark{a}} \\
&  \colhead{NS} 
&  \colhead{BH} 
&  \colhead{Low-Wind} 
&  \colhead{Ang. Mom.} 
}
\startdata

Standard & 88(18) & 53(22) & 76(34) & 91(21) \\
StanCE0.5  & 84(15) & 53(23) & 75(35) & 90(22) \\
StanCE1.0  & 83(15) & 58(24) & 76(34) & 90(23) \\ 
StanCE0.2  & 86(14) & 44(17) & 71(36) & 89(19) \\
IMF2.7  & 89(18) & 57(26) & 77(38) & 92(22) \\
IMF2.7CE0.5 & 85(16) & 60(29) & 79(42) & 91(23) \\
IMF2.7CE1.0 & 84(16) & 62(29) & 68(39) & 91(24) \\
IMF2.7CE0.2 & 87(15) & 48(19) & 73(40) & 90(19) \\
IMF2.7MT1.0 & 82(13) & 55(27) & 76(91) & 91(23) \\
FBBHighWind & 90(19) & 75(1.7) & 28(2.4) & 92(19) \\
FBBLowRad & 0.54(0.20) & 12(6.6) & 92(52) & 27(15) \\
FBBHighRad & 84(6.6) & 33(14) & 35(18) & 79(8.7) \\
Max50  & 87(17) & 51(21) & 58(24) & 91(22) \\
Max100  & 84(18) & 58(30) & 79(43) & 92(27) \\
Max150  & 79(17) & 57(34) & 77(49) & 91(31) \\ 
Max200  & 76(15) & 56(37) & 77(52) & 90(34) \\
Max300  & 69(15) & 54(39) & 75(56) & 89(39) \\
Max400  & 57(13) & 50(39) & 74(60) & 88(47) \\
Max500  & 36(8.4) & 45(38) & 76(66) & 85(57) \\
Max300IMF  & 66(14) & 50(34) & 73(52) & 87(40) \\
MaxHighWind & 87(23) & 79(2.3) & 26(3.3) & 93(24) \\

\enddata 
\tablenotetext{a}{Percentage of systems in each subclass that have  
distances greater than $>10$\,pc ($>100$\,pc)}

\end{deluxetable}

\clearpage

\begin{figure}
\plotone{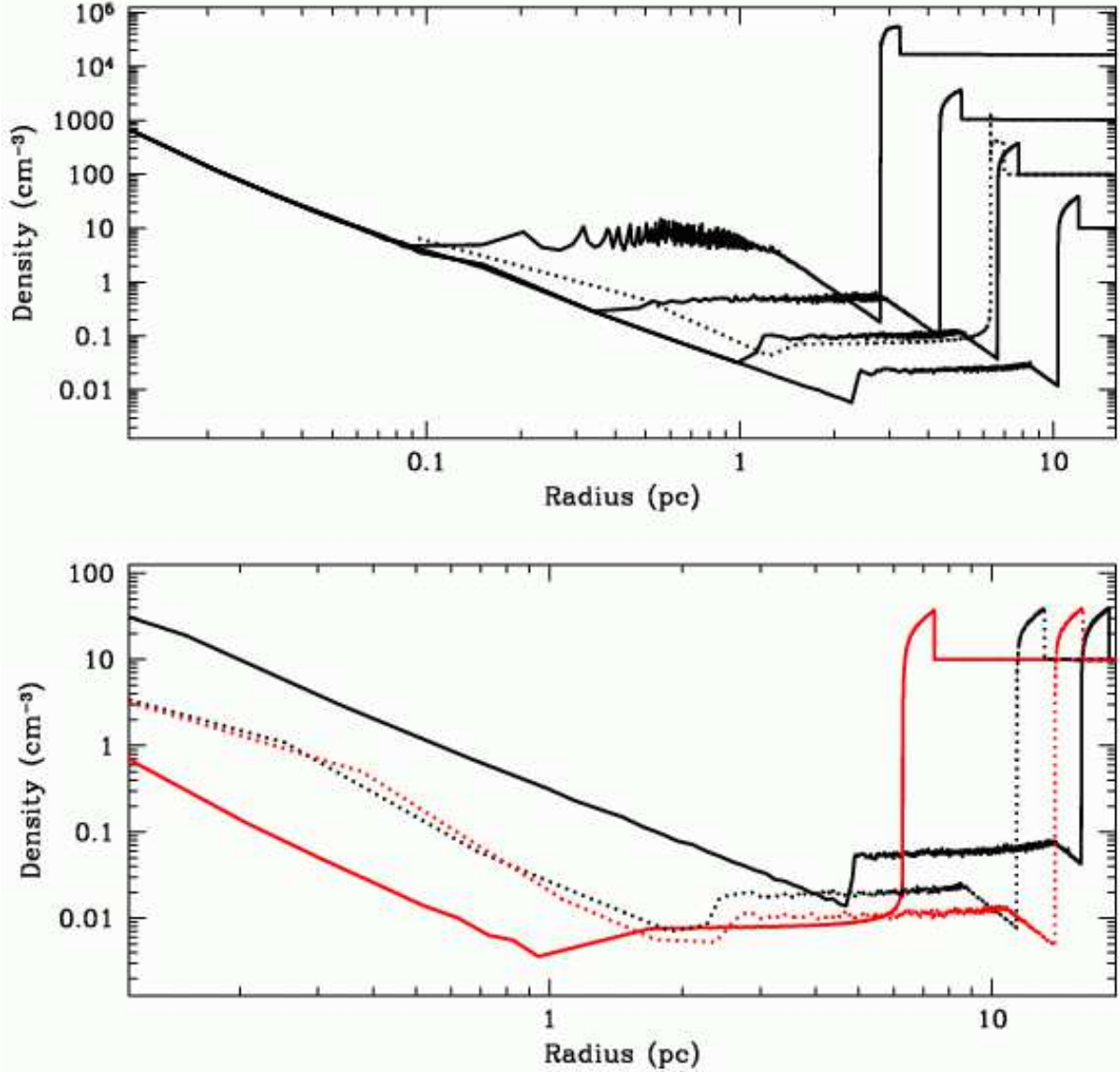}
\caption{Density vs. radius for constant wind models.  The top panel
shows models with mass loss rates of $10^{-5} \, M_\odot y^{-1}$ and
wind velocities of 1,000\,km\,s$^{-1}$ for a range of densities in the
ambient interstellar medium.  The wind extends well into the
interstellar medium (beyond 10\,pc for ambient densities of
10\,cm$^{-3}$).  For afterglow observations, the most critical feature
is the size of the free-streaming region.  It is in this region that
the density follows a $\rho \propto r^{-2}$ dependence on distance.
This region is smaller for higher interstellar densities.  Sound waves
move back and forth through the shocked wind/ISM region, causing
density perturbations, both real and numerical.  The wiggles in the
shocked region of our highest-density cases are numerical in origin.
The dotted line in this top panel shows the result from our simulation
including cooling (see Table 1 for details).  The bottom panel studies
the dependencies on the mass loss rate and the wind velocity around a
model assuming a $10^{-5} \, M_\odot y^{-1}$ mass loss rate, wind
velocity of 1,000\,km\,s$^{-1}$ and ambient density of 10\,cm$^{-3}$.
The solid lines compare different mass loss rates: dark ($10^{-4} \,
M_\odot y^{-1}$), light ($10^{-6} \, M_\odot y^{-1}$).  The dotted
lines correspond to variations in the velocity: dark
(500\,km\,s$^{-1}$), light (2,000\,km\,s$^{-1}$).}
\label{fig:massloss}
\end{figure}
\clearpage

\begin{figure}
\plotone{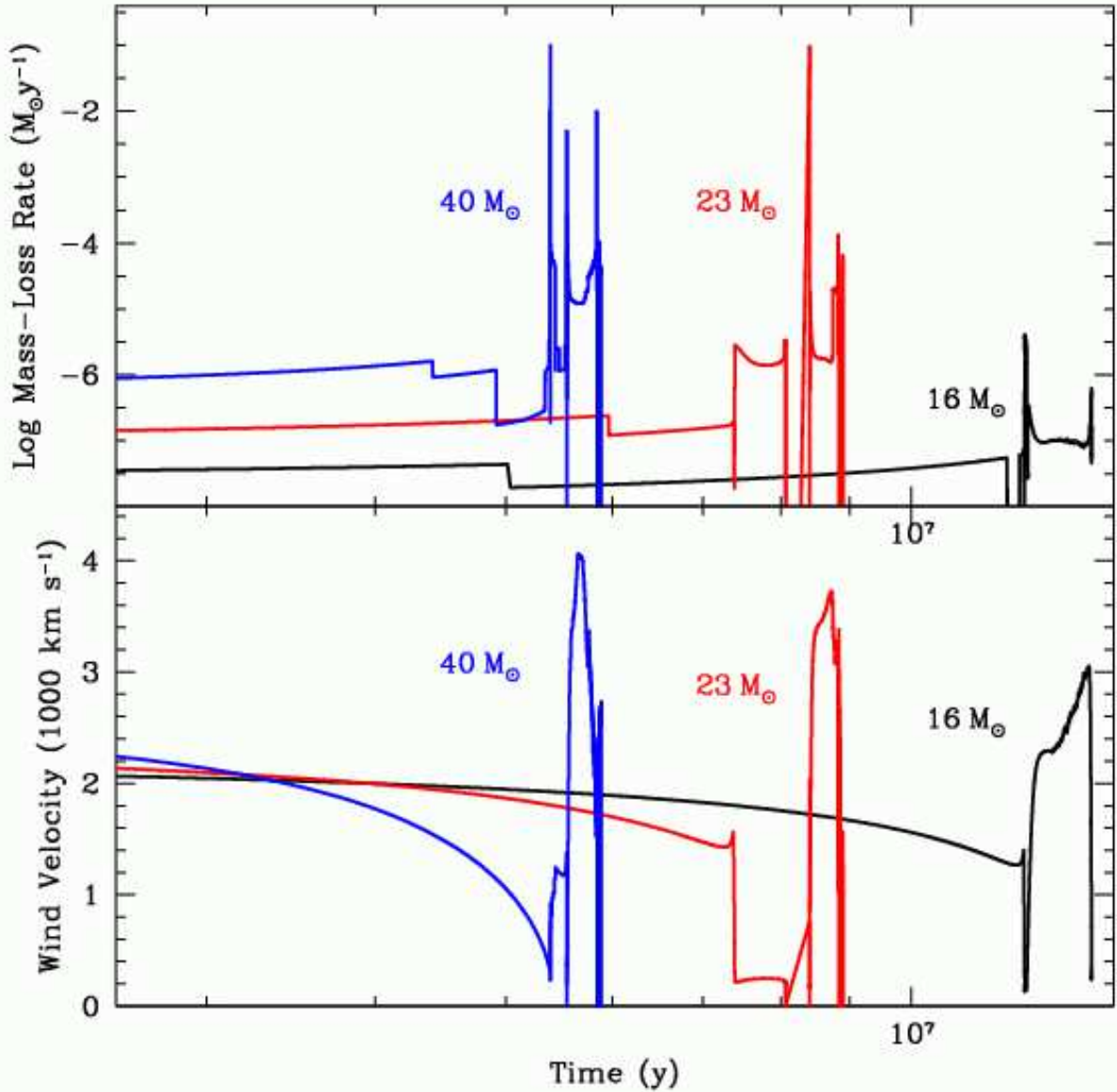}
\caption{Mass loss rate (top) and wind velocity (bottom) time
histories for three stellar models: 16 and 23\,$M_\odot$ binary
systems and a 40\,$M_\odot$ single star.  Note that the mass loss is
far from constant with some pulses lasting for less than 100,000\,y.
The largest mass outflows often are the lowest velocity flows and this
can lead to shocks as the fast moving material following the peak
outflows catches these peaks.}
\label{fig:windparam}
\end{figure}
\clearpage

\begin{figure}
\plotone{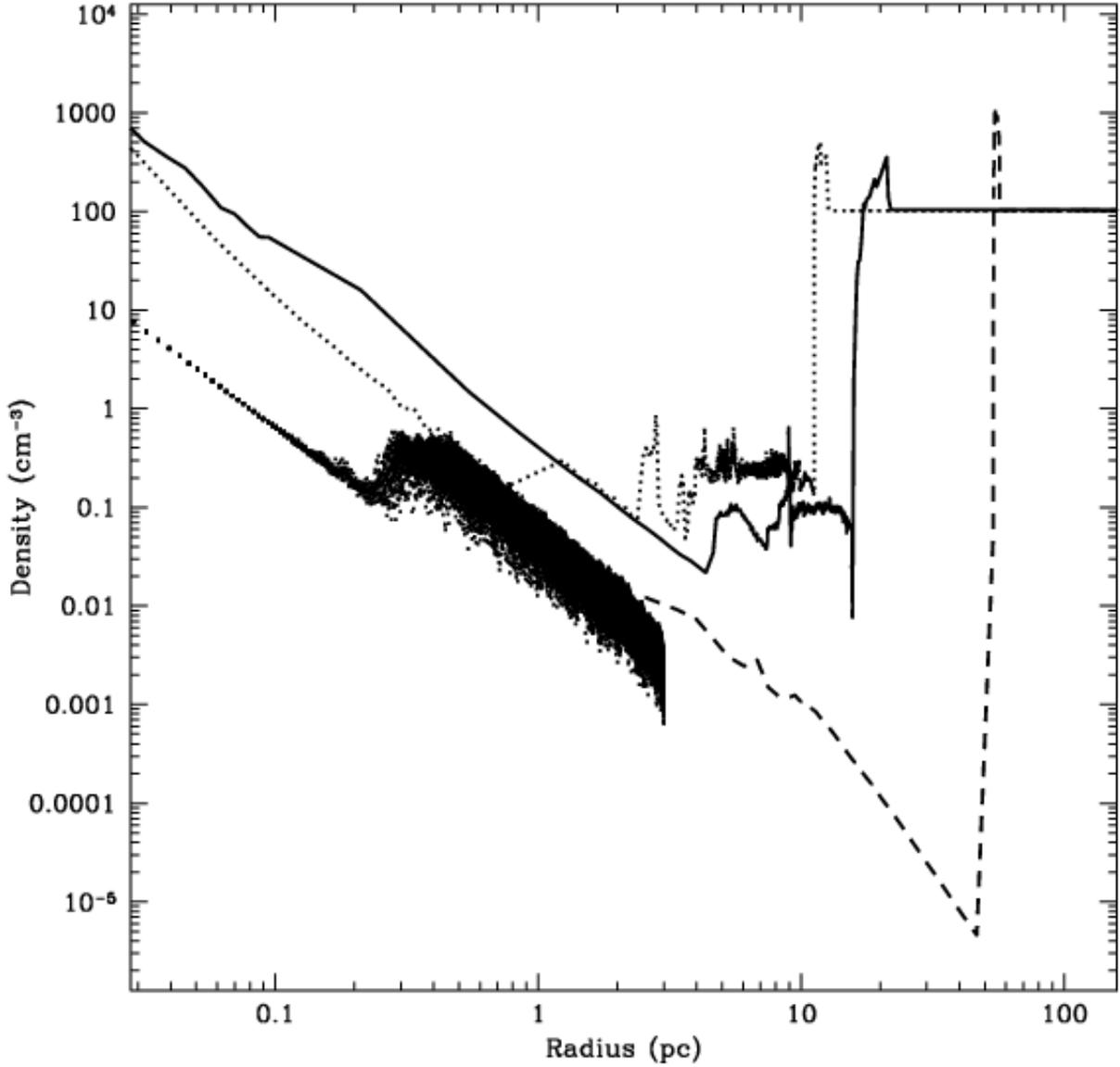}
\caption{Density versus radius for the wind bubbles produced by our 3
stellar models: 40\,$M_\odot$ star (solid line), 23\,$M_\odot$ star
(dotted), 16\,$M_\odot$ star (dashed line).  The effects of the
variable mass loss can be seen (for instance, see the density peak in
the 23\,$M_\odot$ star at roughly 3\,pc), but the density profiles
still follow a roughly $\rho \propto r^{-2}$ profile even beyond the true
termination shock of the free-streaming wind and far enough out that
it should dominate what we see in the GRB afterglow observations.  The
points show the density profile along the z-axis in our 3-dimensional
model.  This model was for the 23\,$M_\odot$ binary star assuming zero
density ambient medium except for a circumstellar disk caused by the
binary interaction (see Fig. 4).}
\label{fig:wind3d-line}
\end{figure}
\clearpage

\begin{figure}
\plotone{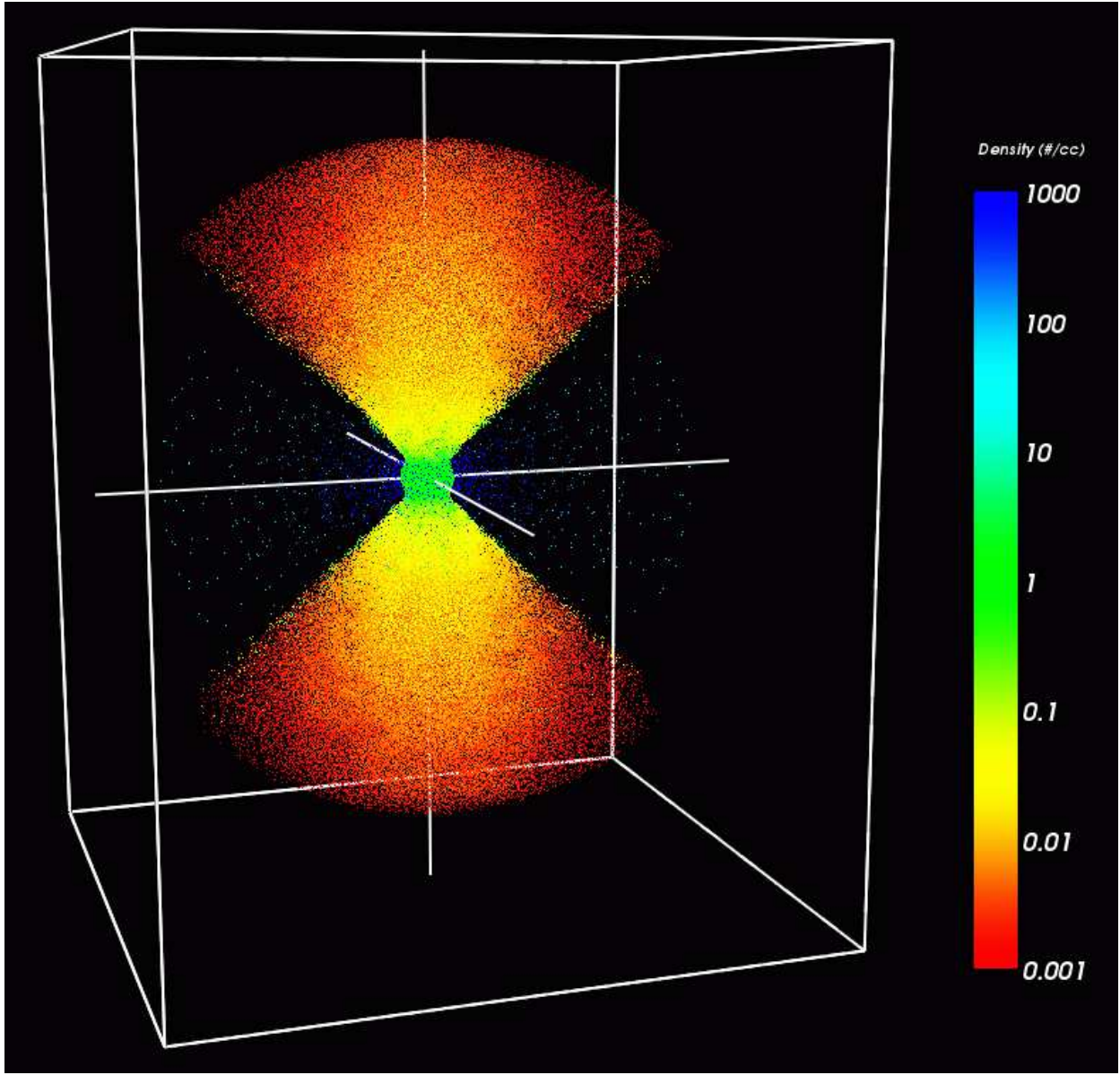}
\caption{A 3-D SPH model of wind from a massive star interacting with a
  wedge of dense material ejected during the common envelope phase of
  a binary system.  The location of each SPH particle is marked by a
  colored point, and the color indicates the number density of gas at
  that location.  The wind-producing star is located at the center of
  the simulation box; the wedge of material ejected during the
  common envelope phase is oriented perpendicular to the $z$
  (vertical) axis.  Wind particles are injected at a radius of
  $0.01$~pc, the inner edge of the dense wedge is located at $0.3$~pc,
  and particles are removed from the simulation when they travel more
  than $3$~pc from the origin.  The density of gas in the wedge is
  much larger than in the subsequently-ejected wind, so the wind is
  obstructed by the wedge and escapes primarily along the $z$ axis.}
\label{fig:wind3d}
\end{figure}
\clearpage

\begin{figure}
\epsscale{.8}
\plotone{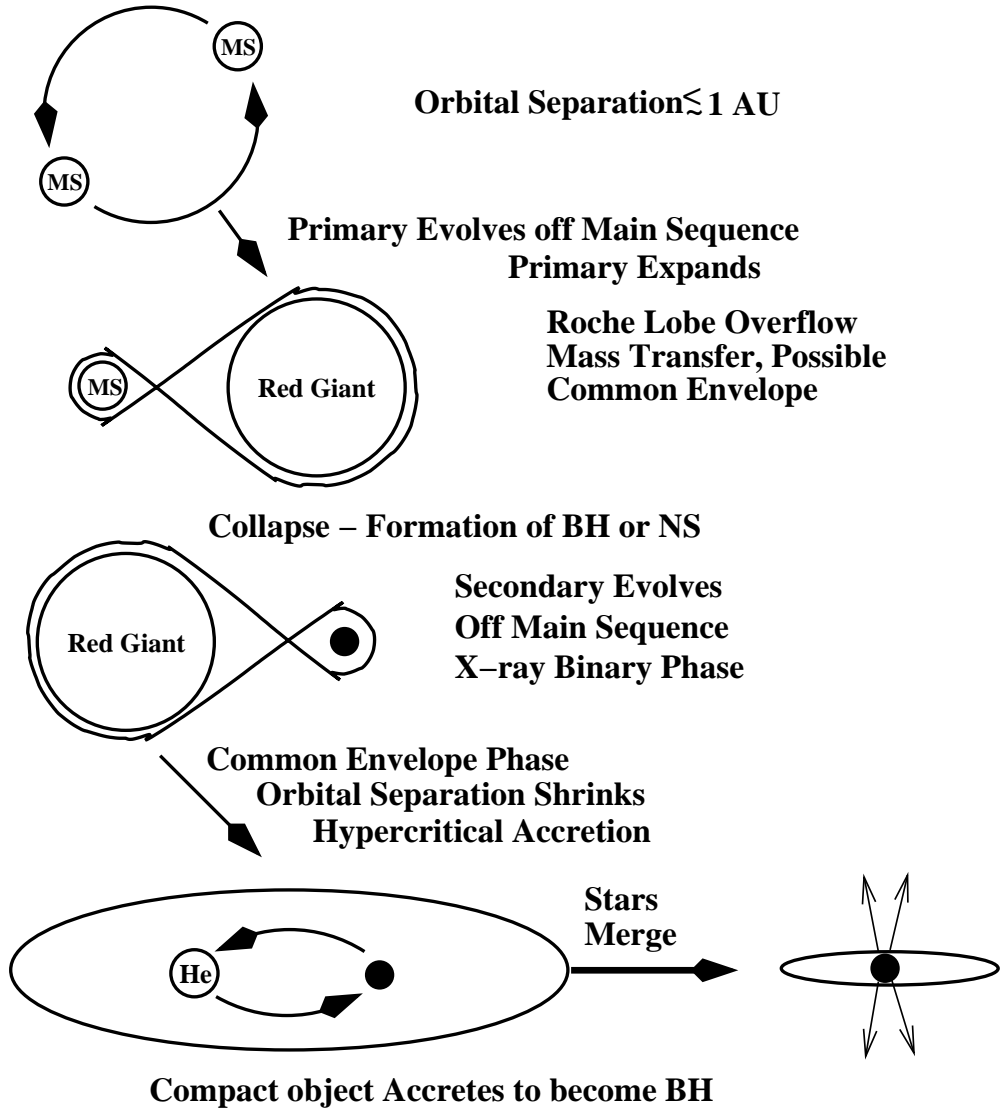}
\epsscale{1}
\caption{Standard formation scenario for He-merger progenitors.  Two
stars in a close binary undergo at least 1 common envelope phase, and
often 2 such phases.  The first, not required, occurs when the primary
(more massive star) evolves off the main sequence.  After the common
envelope phase, the primary collapses to form a neutron star or black 
hole.  If the system remains bound after this collapse and its orbit 
is close enough, the binary will go through a second common envelope 
phase when the companion to the compact remnant evolves off the main 
sequence.  If the compact object merges with its companion, a Helium-merger 
is formed.}
\label{fig:diag1}
\end{figure}
\clearpage

\begin{figure}
\plotone{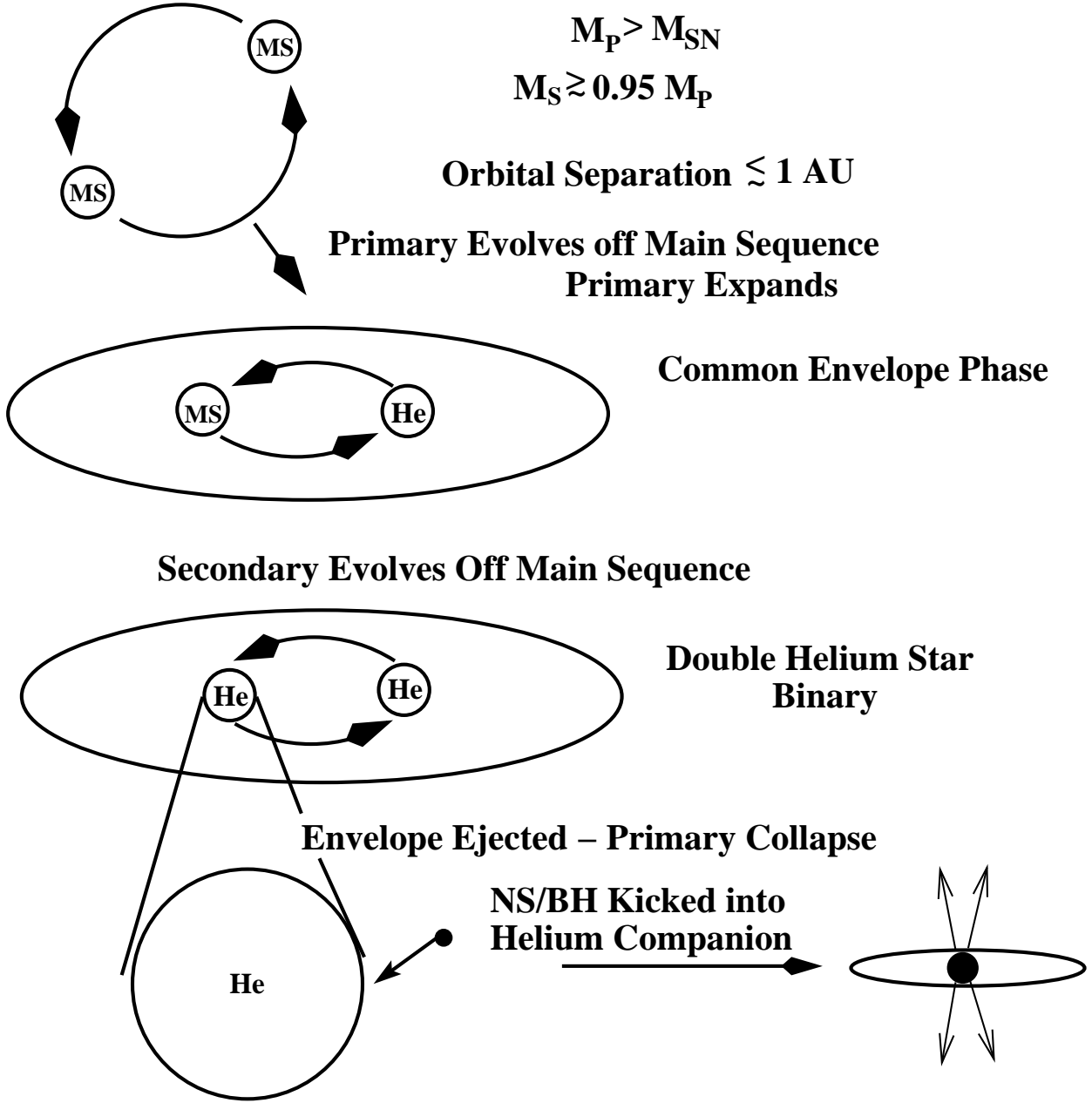}
\caption{Secondary formation scenario of helium-mergers.  Two nearly equal 
mass stars in a close binary evolve off the main sequence together - that is, 
the secondary evolves off the main sequence before the primary collapses.  
The ``joint'' common envelope phase from this evolution produces two helium 
stars.  When one star collapses, its compact remnant may be kicked into 
the helium companion, producing a helium-merger.}
\label{fig:diag2}
\end{figure}
\clearpage

\begin{figure}
\plotone{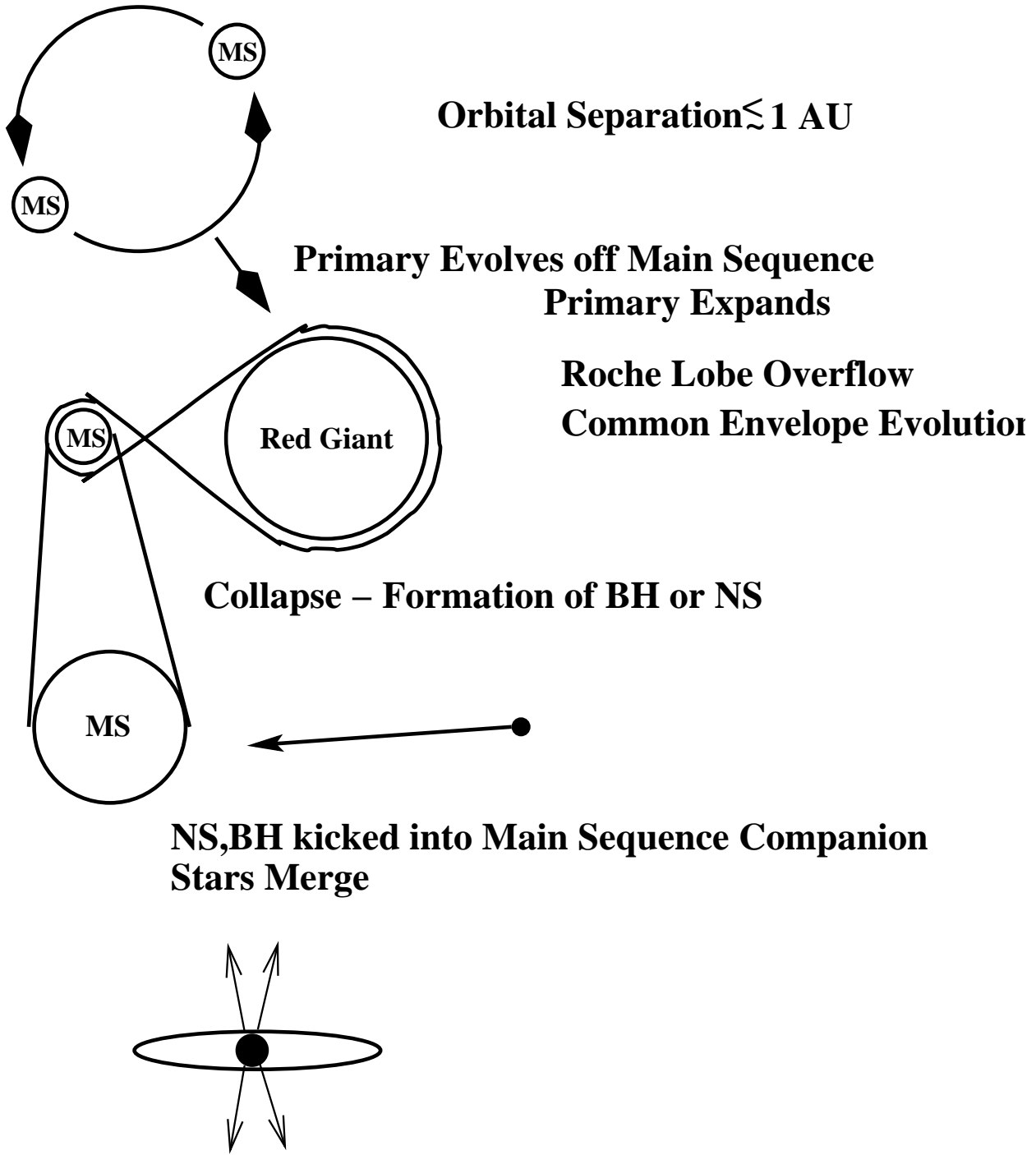}
\caption{Alternate scenario for producing merged systems.  In this close 
binary system, the collapse of the primary star kicks the compact remnant 
into its main-sequence companion.  Although not strictly a Helium-merger 
because the companion is still on the main sequence (and hence not quite 
as dense and less likely to be an ideal GRB progenitor), we include it 
in our study.}
\label{fig:diag3}
\end{figure}
\clearpage

\begin{figure}
\caption{A series of plots showing the population for all our
He-merger formation scenarios (either total number or fraction) for
our Standard model versus system velocity, merger time or distance
travelled.  The different curves denote different subclasses and have
the same definitions for each plot: dark solid - all systems, dark
dotted - angular momentum, light solid - BH Collapse, light dotted -
NS Collapse, and dashed - Low-Wind (see Table 3 for details).  a)
Number of binaries per 1\,km\,s$^{-1}$ bin.  The peak in low velocity
systems only occurs in the neutron star compact object and small
companion mass subclasses.  The velocities are only high enough to
escape the host galaxy gravitational potential if the host is an
extremely low-mass galaxy.  Note that the peak in low-velocity 
systems is dominated by the NS Collapse and Low-Wind models.  It 
is difficult for those systems that form neutron stars to remain bound, 
requiring the kick to counter the momentum lost in the 
supernova ejecta.  The systems that are more likely to remain bound 
for these NS binaries are those where the countering effects lead to 
small proper motions.  b) Number of binaries per log Time bin (900
bins in plot).  The large spike in binaries with merger times
$\lesssim 10$y arises from the Brown and MS Merger formation scenarios.
These scenarios only produce immediate mergers.  The bulk of long
merger times are produced by those mergers with low wind (and hence
low-mass) companions.  c) Number of binaries per 1\,pc bin.  The spike at low
distances corresponds to the Brown and MS-merger formation scenarios where the
merger time is nearly immediate (see figure 8b).  Even for the primary 
He-merger scenario, most mergers occur within 100\,pc of ther formation site.
A small fraction occur beyond 1\,kpc (Table 6).  d-f) Same as figures a-c) 
but showing the relative fractional distribution (Number in a bin divided by 
the total number of a given subclass).}
\label{fig:nvsv1}
\end{figure}
\epsscale{.8}
\clearpage
\plottwo{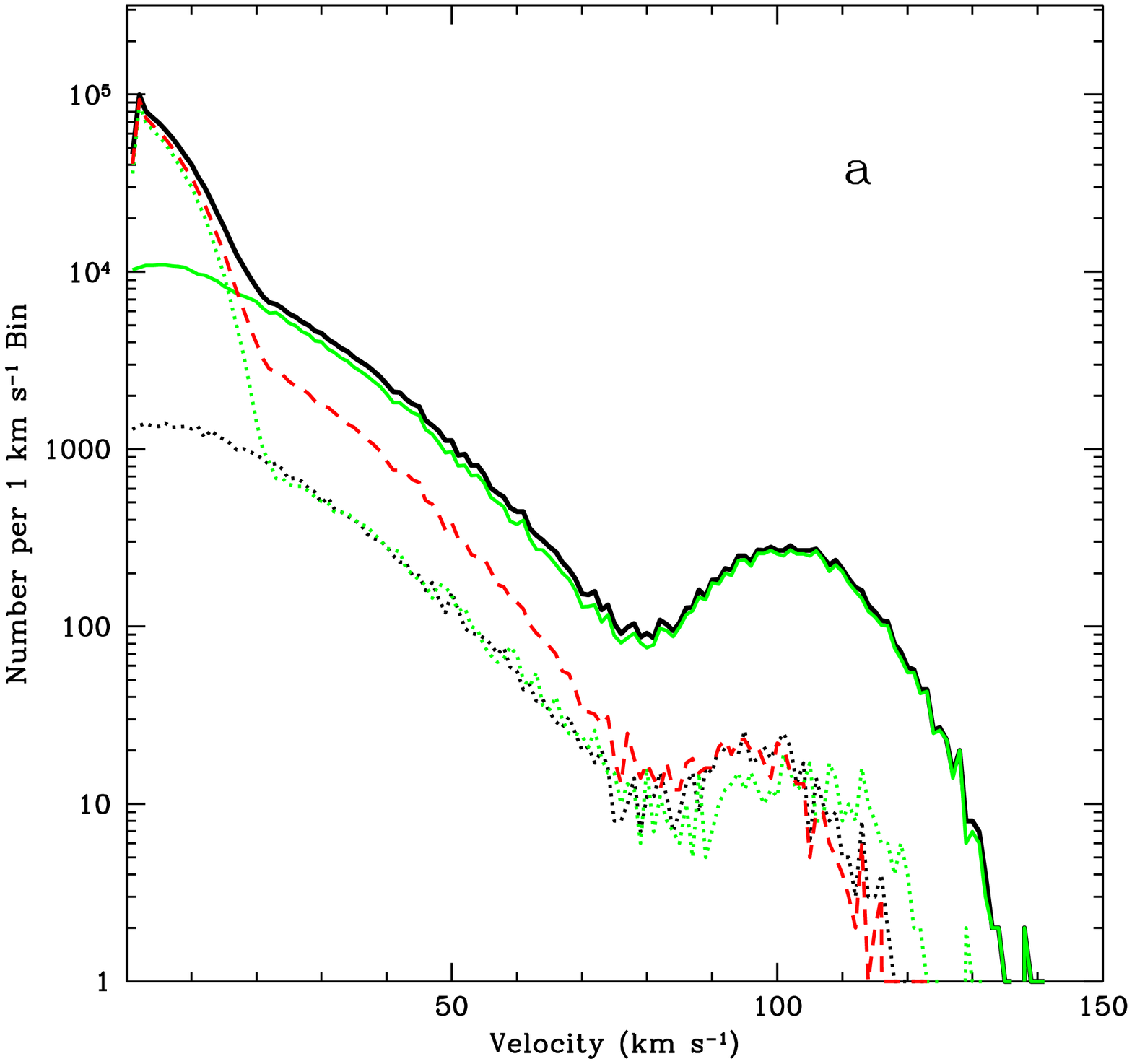}{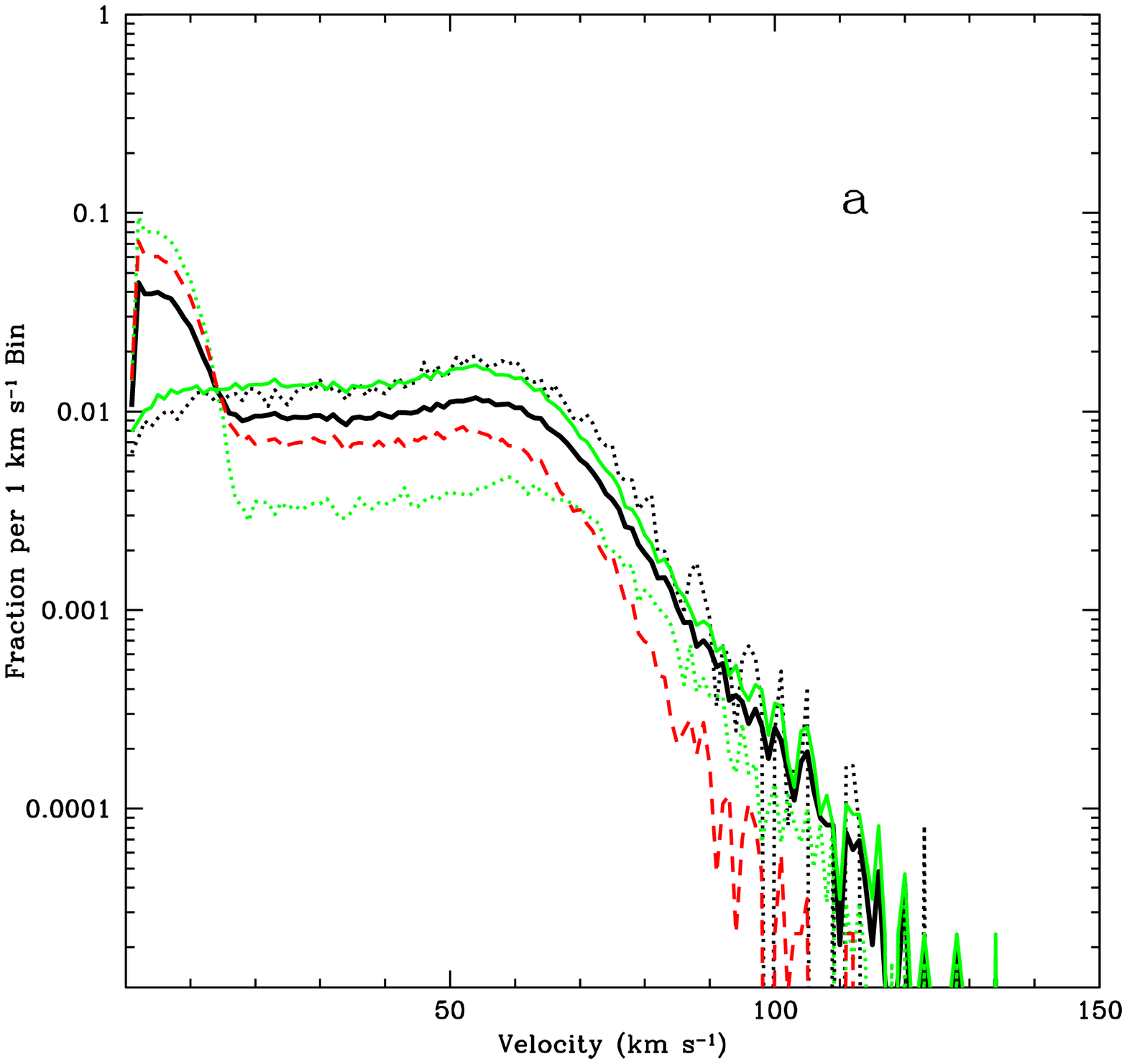}
\plottwo{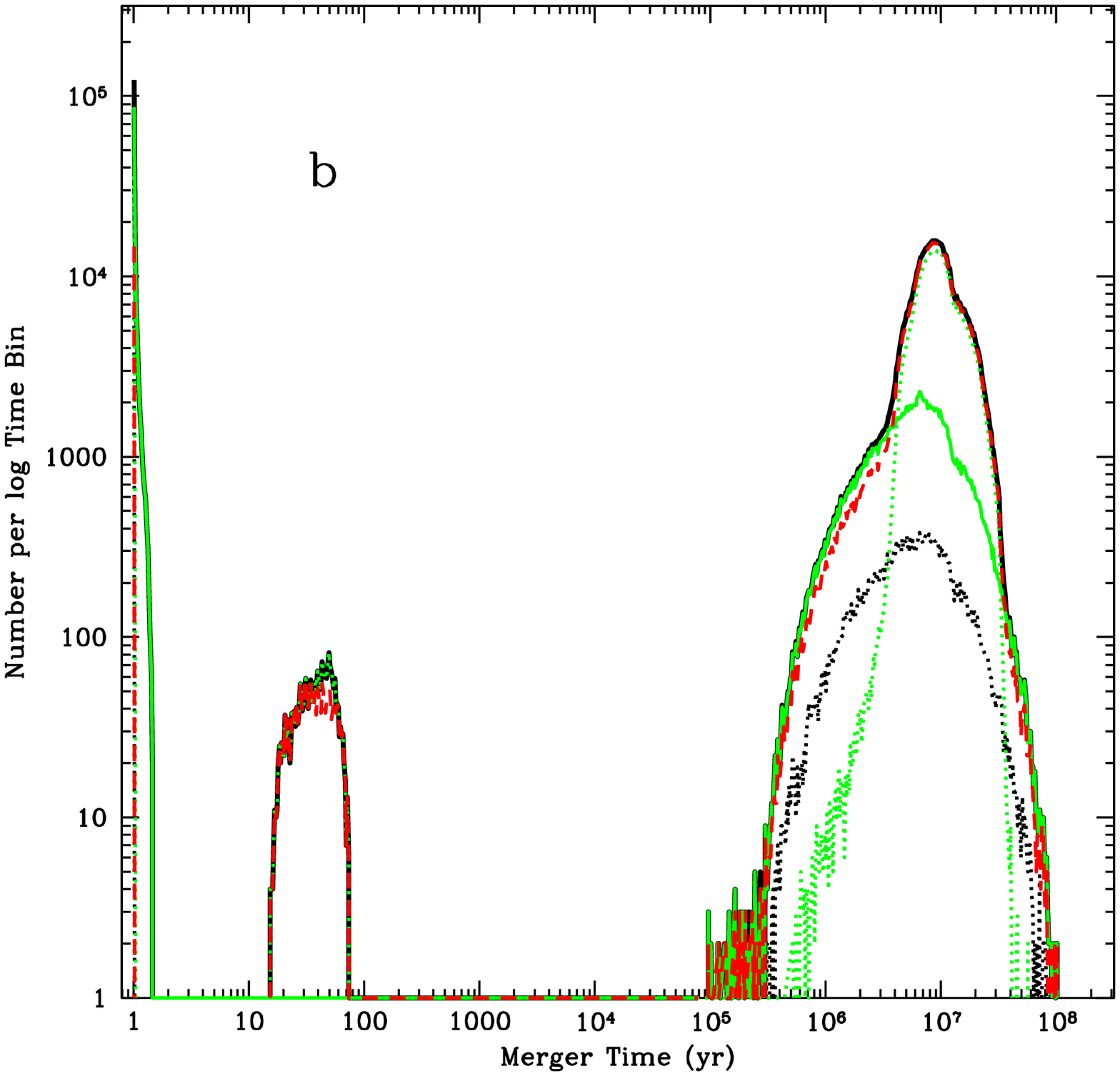}{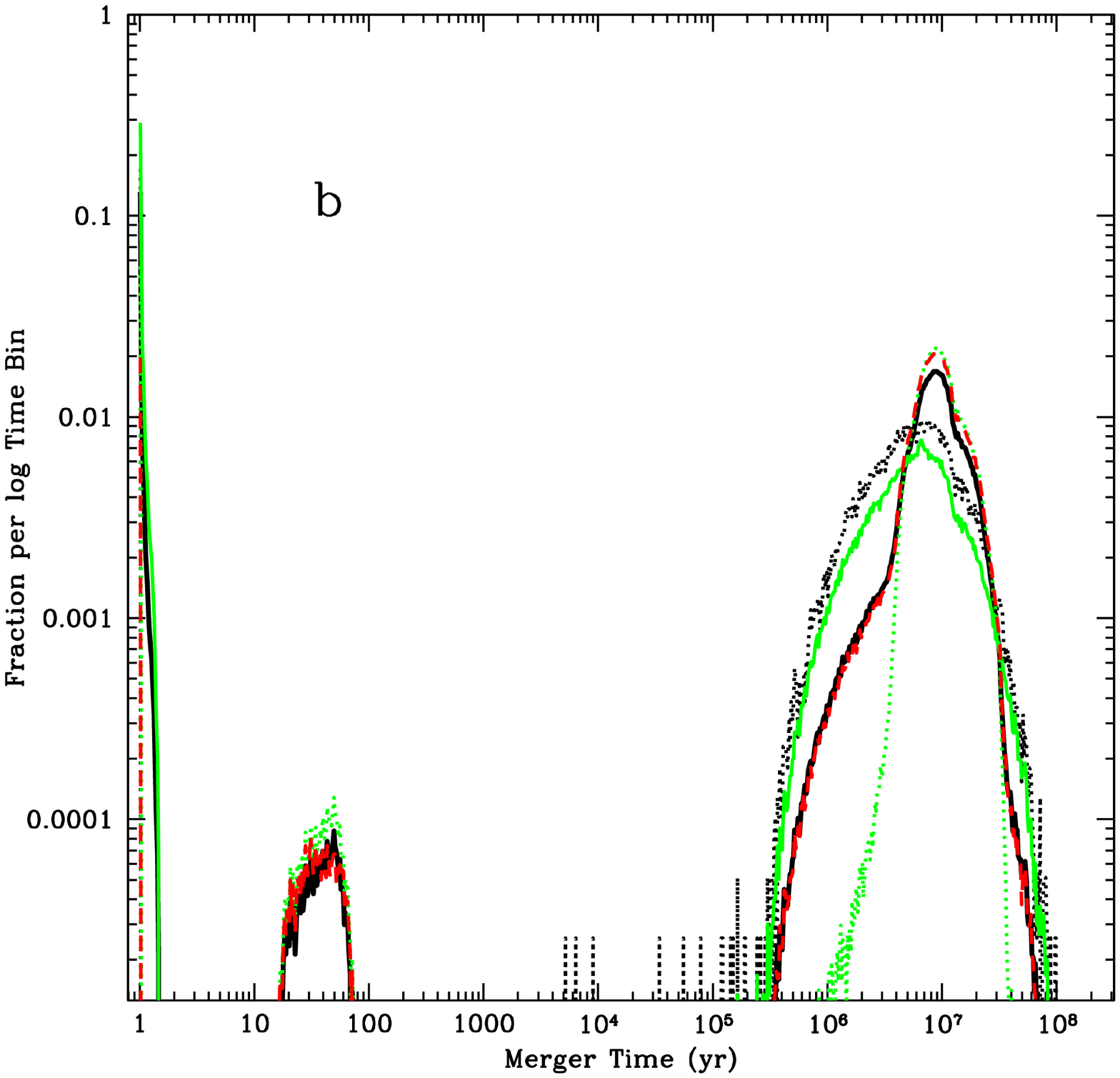}
\plottwo{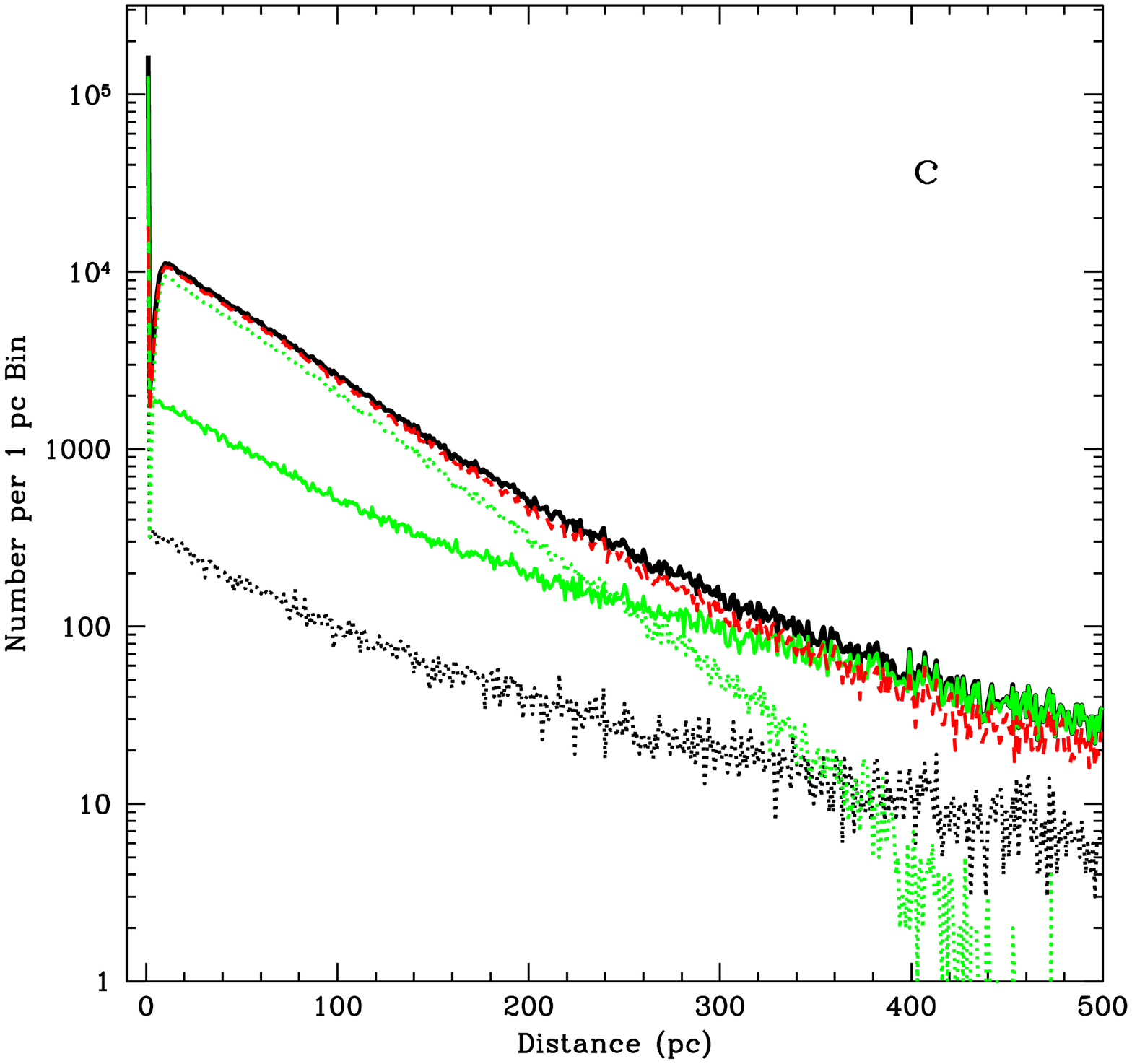}{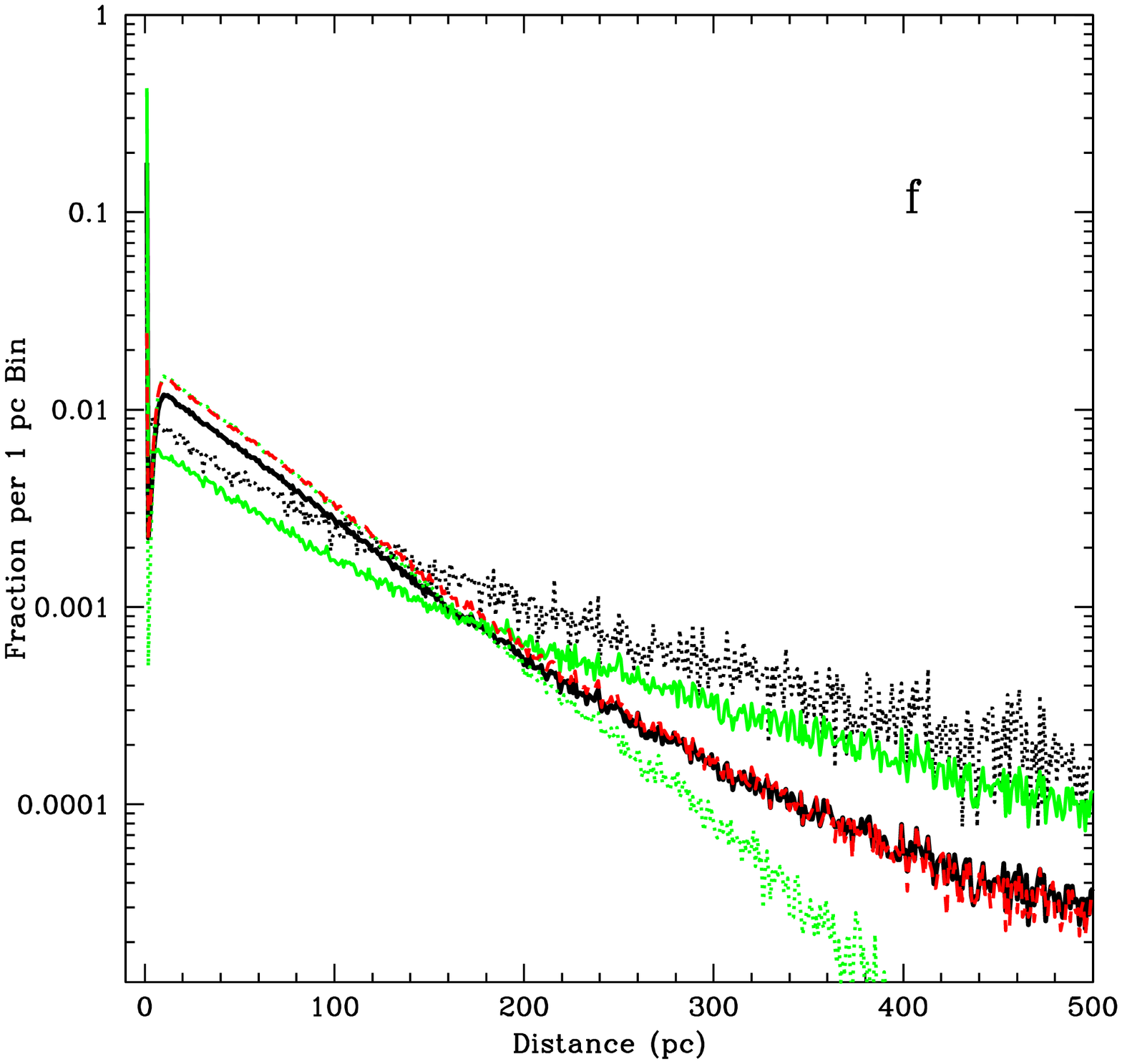}
\centerline{Fig. 8. ---}
\clearpage
\begin{figure}
\caption{Relative fractional distributions for model Max300.  These 
3 figures are identical in scale to figures 8d-f and can be compared
directly to the results of the standard model in figure 8.  Note that
because the kick in this model is single-peaked, the double peak in
the velocity distribution does not occur in this model.}
\label{fig:dist-max}
\end{figure}
\clearpage
\plottwo{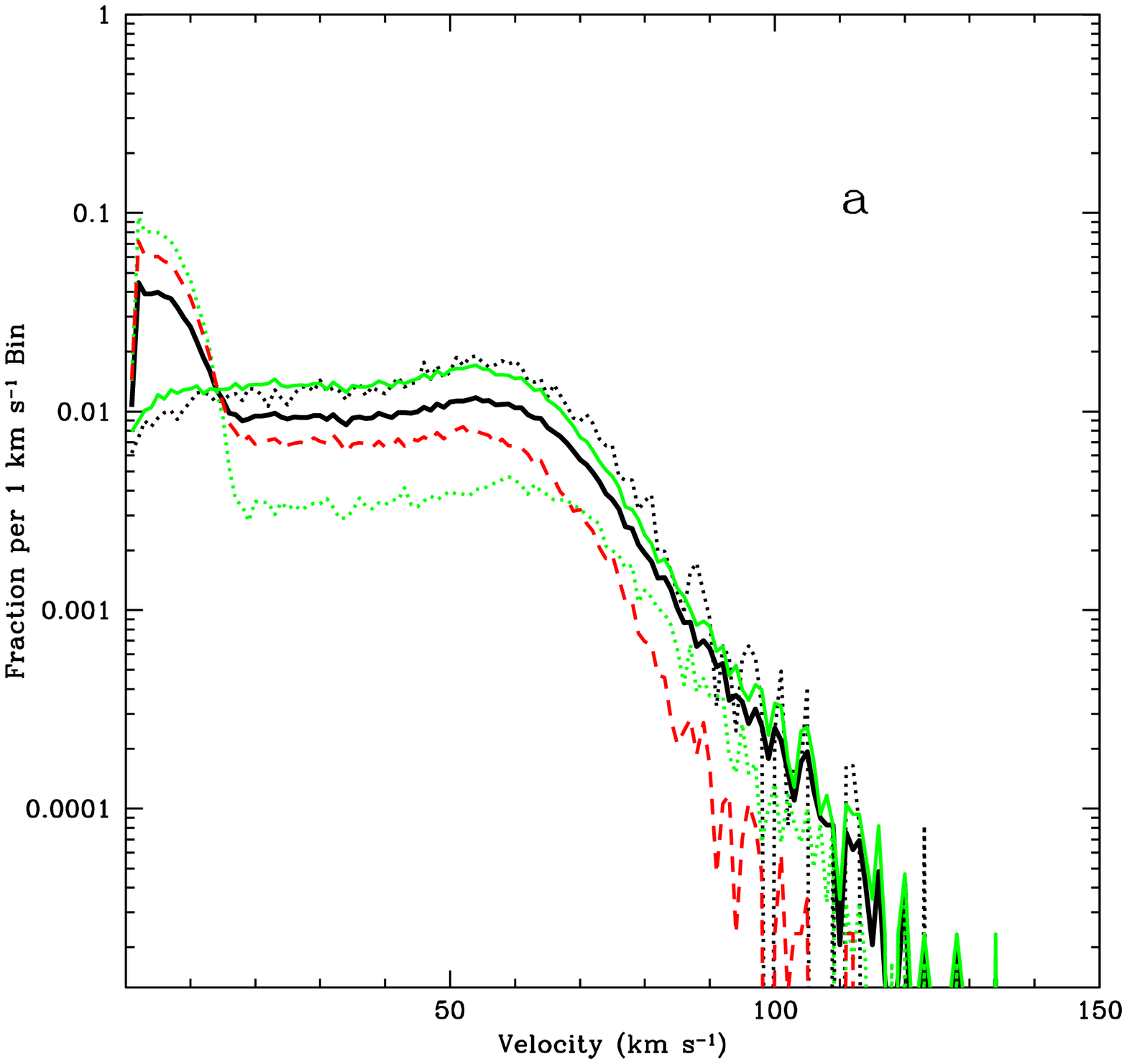}{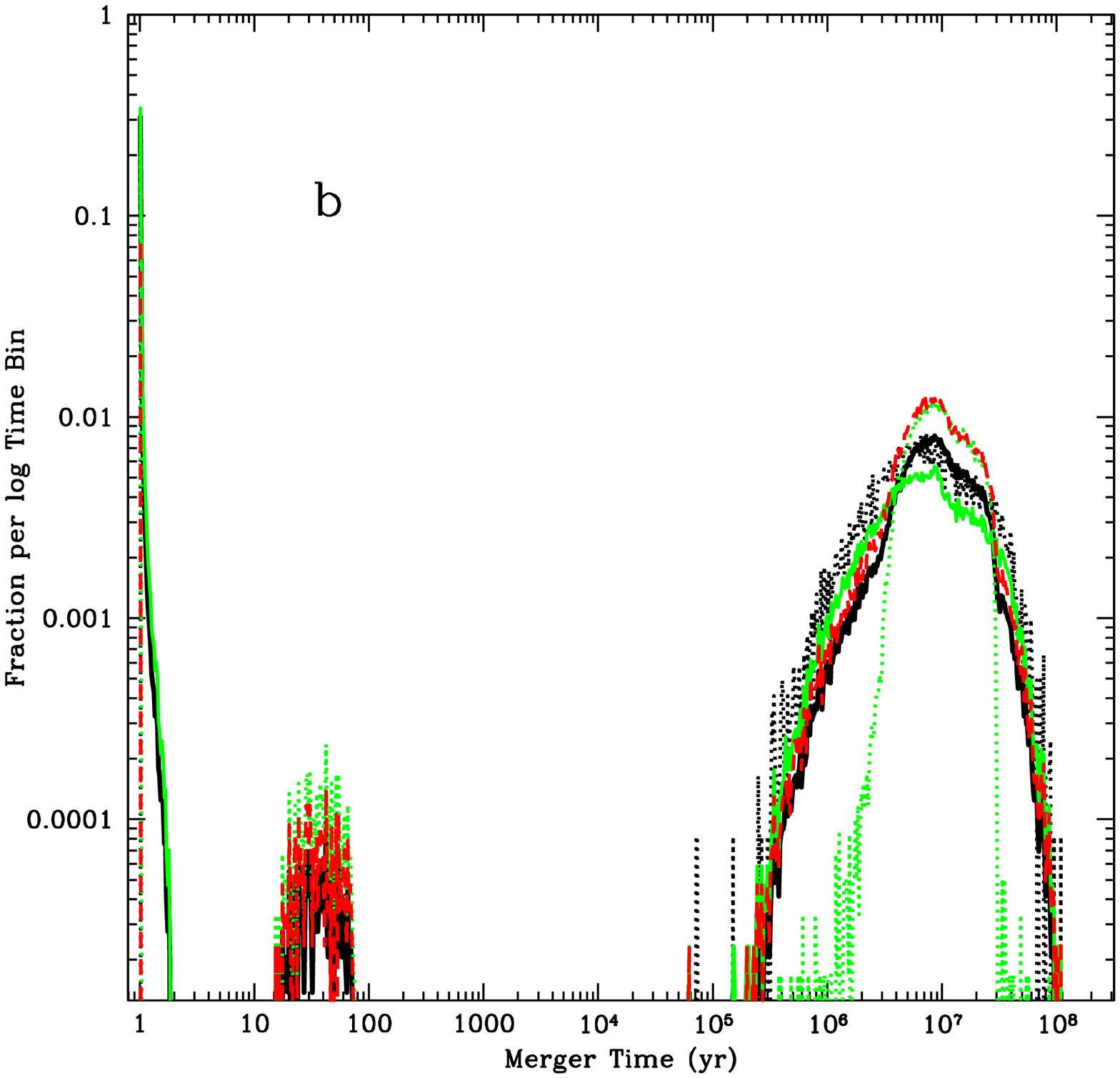}
\plotone{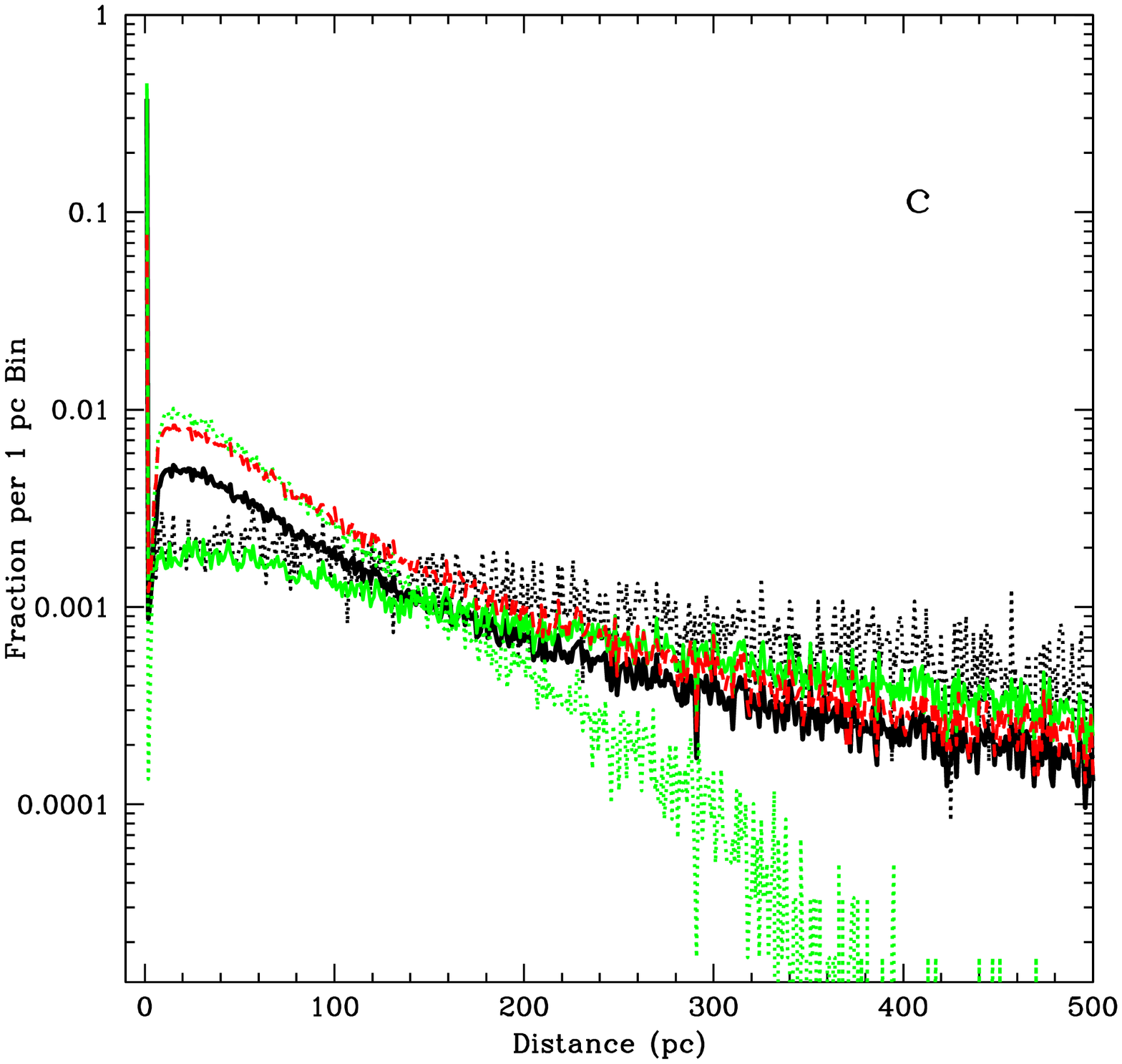}{}
\centerline{Fig. 9. ---}
\clearpage
\begin{figure}
\caption{Same as Figure 9 showing just the total He-merger
distributions for a range of population synthesis progenitors: black
solid (Standard), red solid (FBBLowRad), red dotted (FBBHighWind),
green solid (StanCE1.0), green dotted (StanCE0.2), blue solid
(IMF2.7MT1.0), blue dotted (IMF2.7CE0.2), cyan solid (IMF2.7CE0.5).
The largest variations are caused by those models where the stellar
radii or stellar mass loss was modified.}
\label{fig:mcomp}
\end{figure}
\clearpage
\plottwo{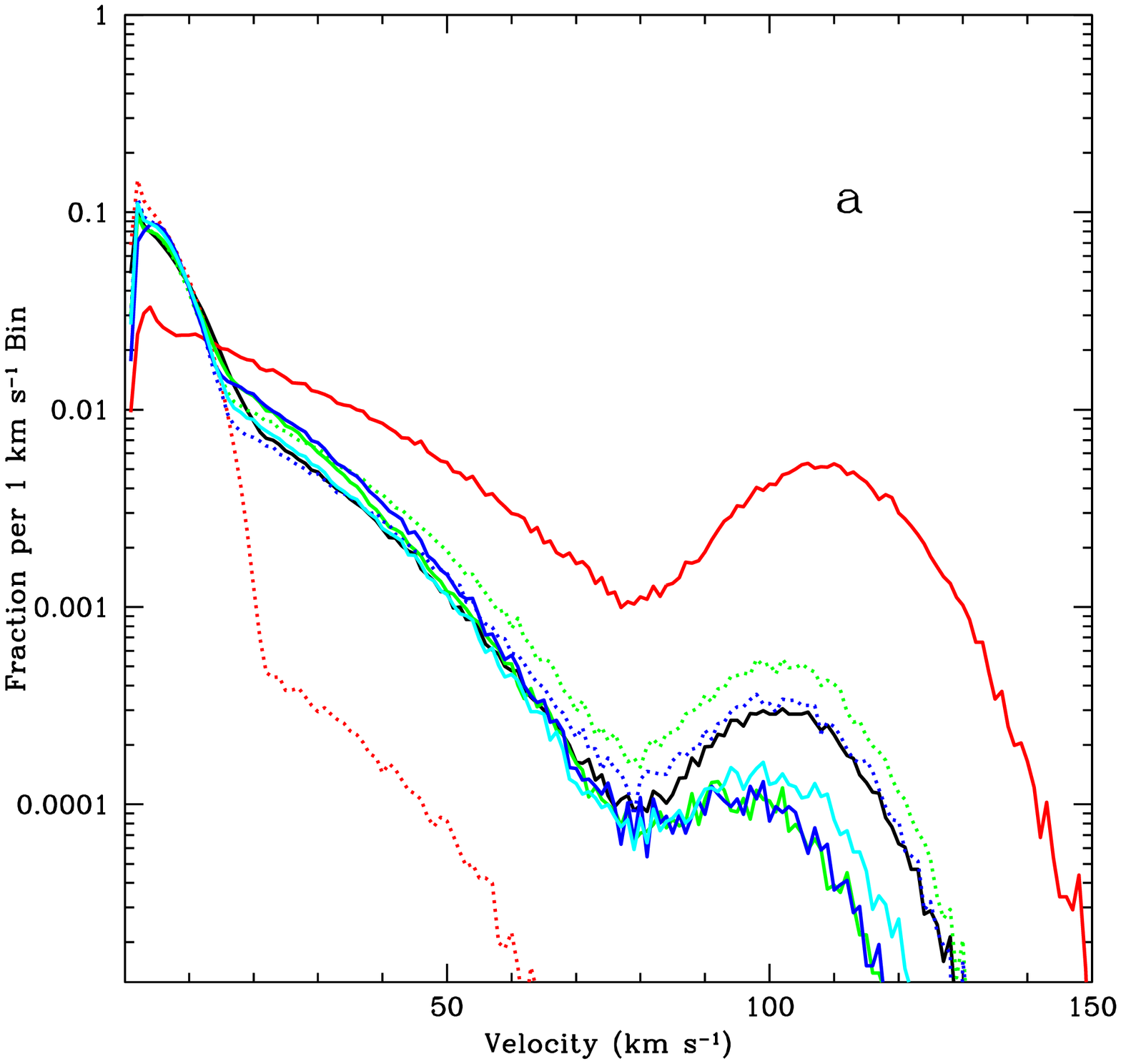}{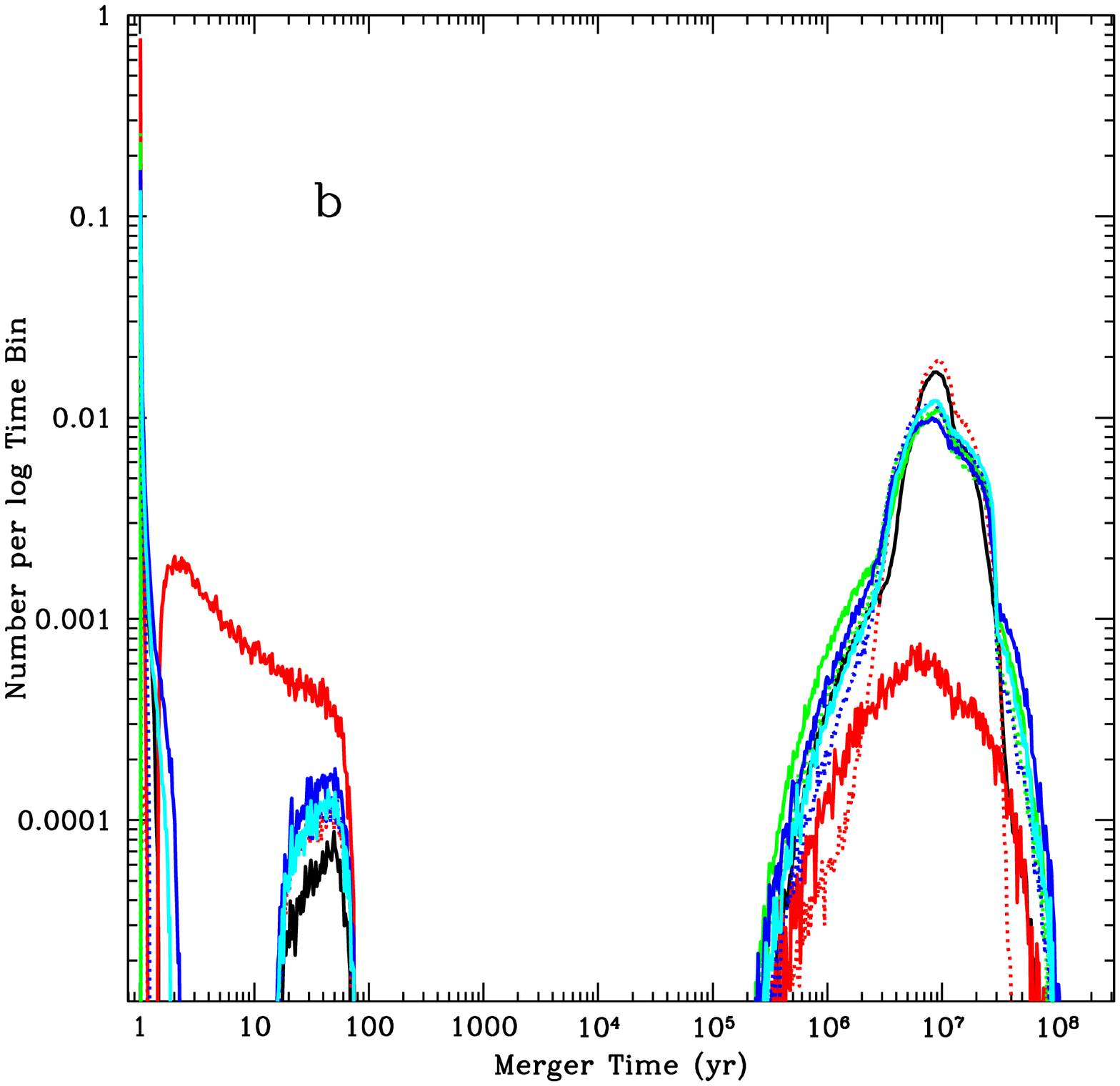}
\plotone{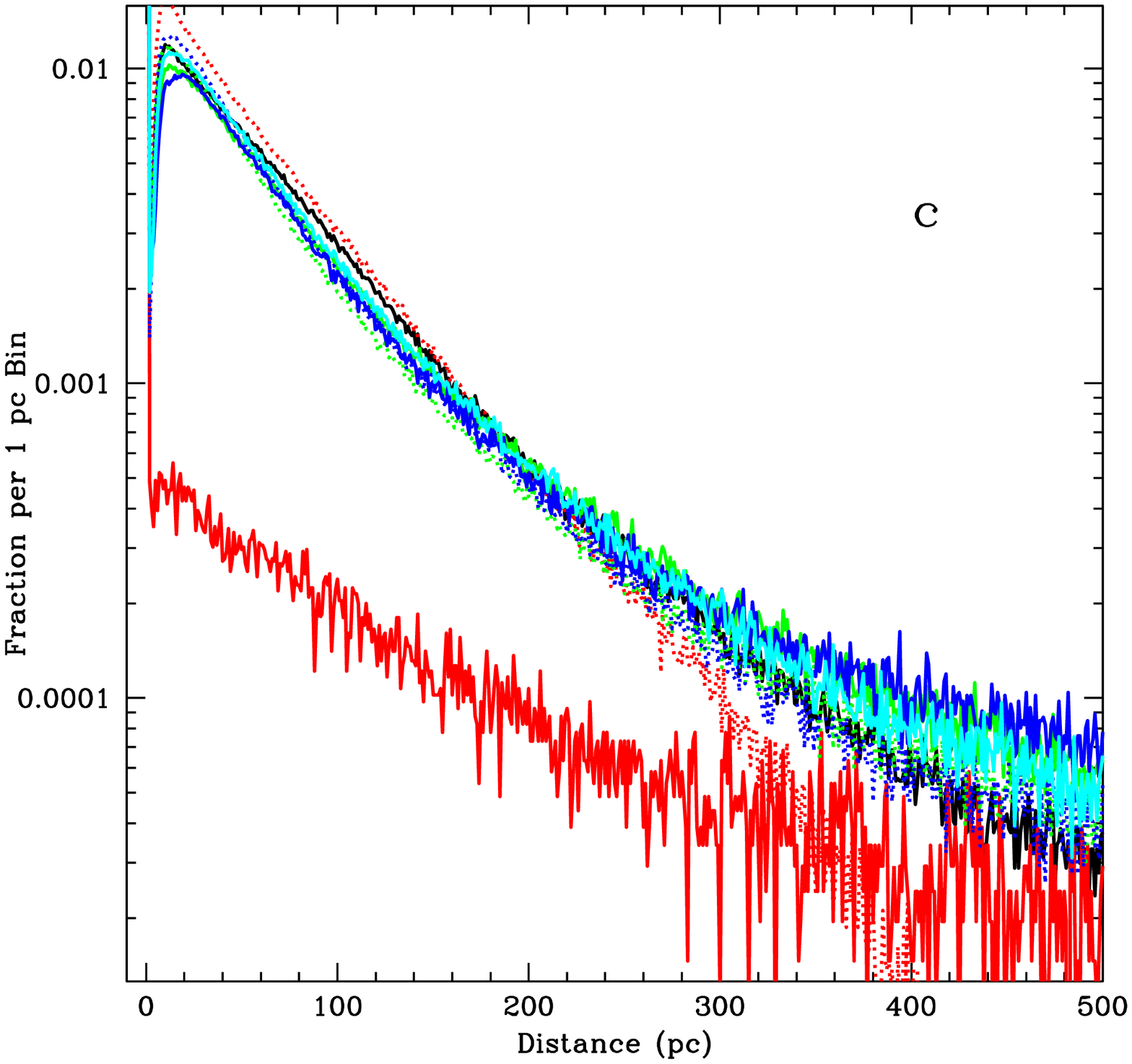}{}
\centerline{Fig. 10. ---}
\clearpage
\begin{figure}
\caption{Same as Figure 9 showing just the total He-merger
distributions as a function of kick magnitude (using single peaked 
maxwellian distributions): blue solid (Max100), cyan solid (MAX150), 
green solid (MAX200), blue dotted (MAX300), cyan dotted (MAX400), green dotted (MAX500),
red dotted (MAX300IMF).  Note that the kick 
distribution has a much larger effect than changes in the initial 
mass function.}
\label{fig:vcomp}
\end{figure}
\clearpage
\plottwo{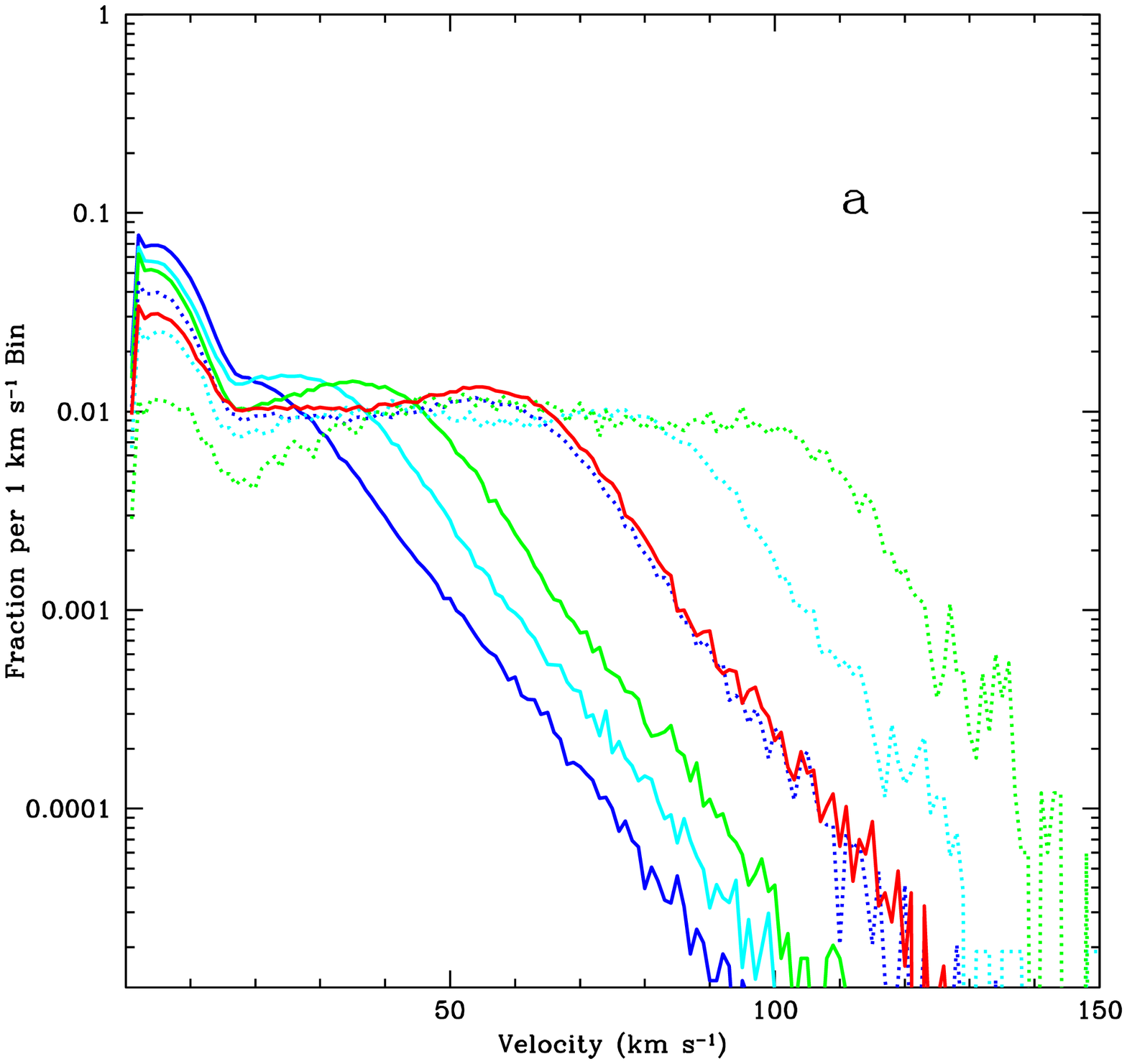}{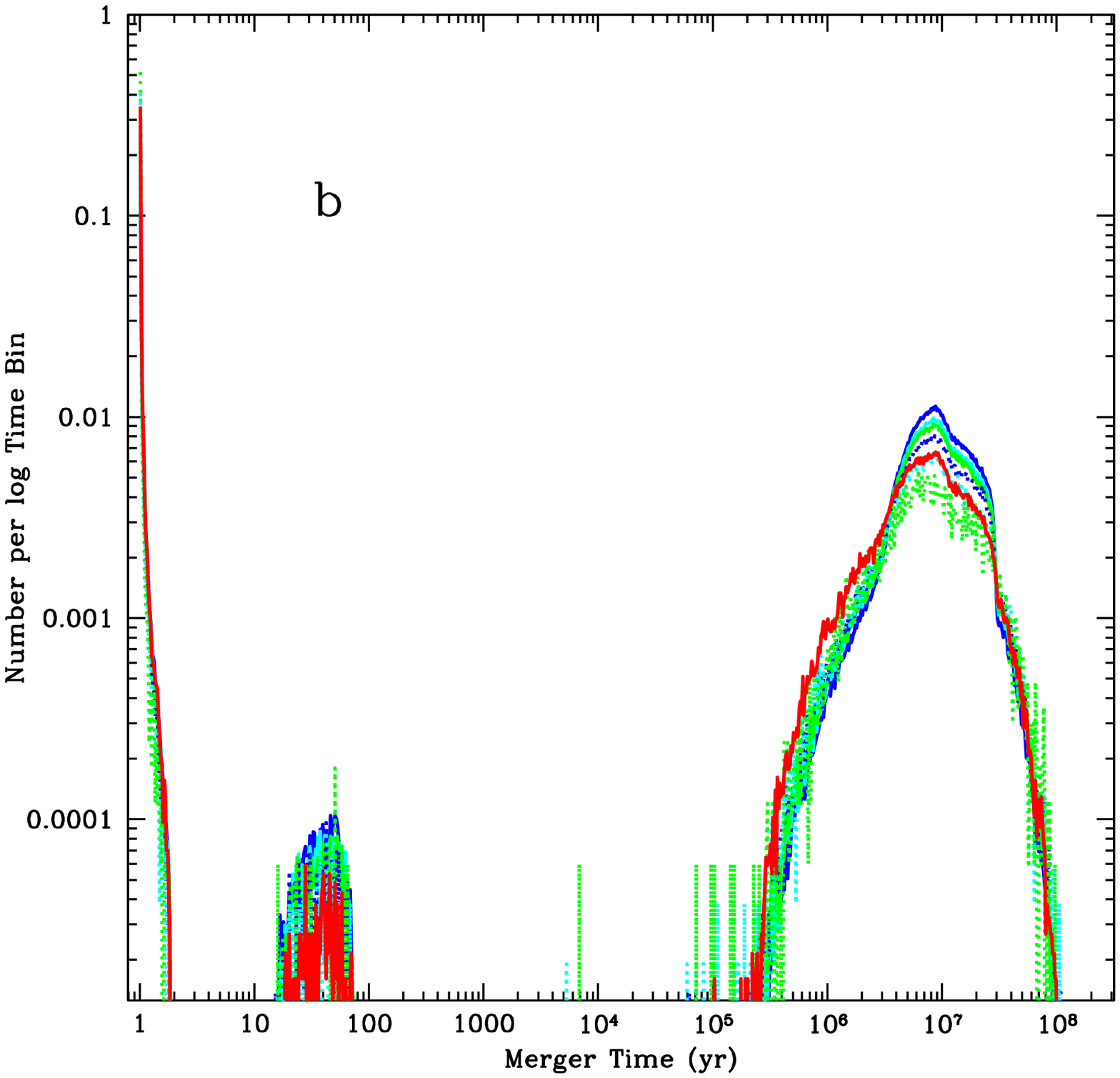}
\plotone{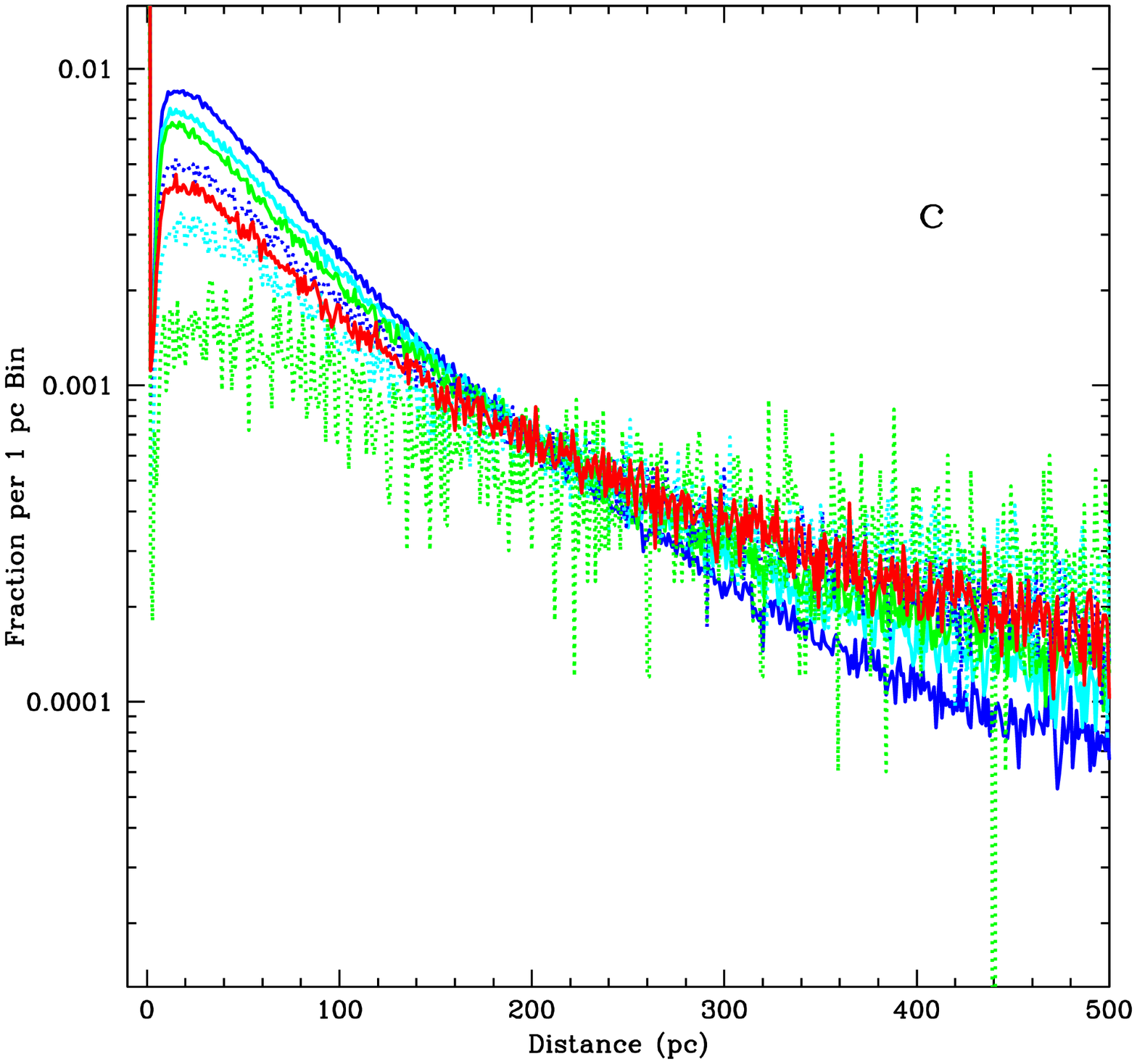}{}
\centerline{Fig. 11. ---}

\end{document}